\documentclass[journal,10pt]{IEEEtran}
\pdfoutput=1
\usepackage{cite}
\usepackage{amsfonts}
\usepackage{amssymb}
\usepackage{amsmath}

\usepackage{mathrsfs}
\usepackage{stfloats}
\usepackage{bm}
\usepackage{graphicx}
\usepackage{subfigure}
\usepackage{color}
\usepackage{algorithm}
\usepackage{algorithmic}
\IEEEaftertitletext{\vspace{-1.5\baselineskip}}

\newtheorem{theorem}{Theorem}

\newtheorem{remark}{Remark}

\begin{document}
%
\title{Spectral Efficiency of Unicast and Multigroup Multicast
Transmission in Cell-free Distributed Massive MIMO Systems}
%
%
%

\author{Jiamin~Li,~\IEEEmembership{Member,~IEEE,}
	~Qijun~Pan,~\IEEEmembership{Student Member,~IEEE,}
	~Zhenggang~Wu,~\IEEEmembership{Student Member,~IEEE,}
	~Pengcheng~Zhu,~Dongming~Wang,~\IEEEmembership{Member,~IEEE,}
		and~Xiaohu~You,~\IEEEmembership{Fellow,~IEEE}
\thanks{\emph{Corresponding author: Pengcheng Zhu (e-mail: p.zhu@seu.edu.cn)}}
\thanks{This work was supported in part by the National Natural Science Foundation of China (NSFC) under Grant 61971127, 61871465, 61871122, by the National Key Research and Development Program under Grant 2020YFB1806600, and by Natural Science Foundation of Jiangsu Province under Grant BK20180011.}
\thanks{J. Li, D. Wang and X. You are with the National Mobile Communications 
Research Laboratory, Southeast University, Nanjing 210096, China, and also with 
Purple Mountain Laboratories, Nanjing 211111, China (e-mail: jiaminli, wangdm, 
xhyu@seu.edu.cn).}
\thanks{Q. Pan, Z. Wu and P. Zhu are with National Mobile Communications 
Research Laboratory, Southeast University, Nanjing 210096, China (e-mail: 
panqijun,220200712, p.zhu@seu.edu.cn).}

}

%
%



\maketitle

\begin{abstract}
In this paper, we consider a joint unicast and multi-group
multicast cell-free distributed massive multiple-input multiple-output (MIMO)
system, while accounting for co-pilot assignment strategy based channel
estimation, pilot contamination and different precoding schemes. Under the
co-pilot assignment strategy, we derive the minimum-mean-square error (MMSE)
channel state information (CSI) estimation for unicast and multicast users.
Given the acquired CSI, the closed-form expressions for downlink achievable
rates with maximum ratio
transmission (MRT), zero-forcing (ZF) and MMSE beamforming are derived. Based on these expressions, we propose an efficient power allocation scheme by solving a multi-objective
optimization problem (MOOP) between maximizing the minimum spectral efficiency
(SE) of multicast users and maximizing the average SE of unicast users with
non-dominated sorting genetic algorithm II (NSGA-II).
Moreover, the MOOP is converted into a deep learning (DL) problem and solved by
an unsupervised learning method to further promote computational efficiency.
Numerical results verify the accuracy of the derived closed-form expressions
and the effectiveness of the joint unicast and multigroup multicast
transmission scheme in cell-free distributed massive MIMO systems. The SE
analysis under various system parameters and the trade-off regions between
these two conflicting optimization objectives offers numerous flexibilities for
system optimization.

\end{abstract}




%
\IEEEpeerreviewmaketitle

\section{Introduction}
%
%
%
%

\IEEEPARstart{C}{ell-free}  distributed massive multiple-input multiple-output
(MIMO) systems are practical and scalable scenarios of MIMO network
\cite{nayebi2015cell,buzzi2019user,ngo2015cell}. By reaping the benefits from
both massive MIMO and network MIMO systems, cell-free distributed massive MIMO
can effectively improve the spectral efficiency (SE), energy efficiency (EE)
and reliable data transmission
\cite{li2017downlink,parida2018downlink,riera2019decentralization}.
In cell-free distributed massive MIMO systems, connected to a central
processing unit (CPU), a great number of remote antenna units (RAUs) are
geographically distributed and coherently serve the users.
Compared with centralized massive MIMO, cell-free distributed
massive MIMO provides a higher macro-diversity gain and lower proximity, thus
achieves better performance
\cite{yaacoub2016overview,	larsson2016joint	}.
{ \cite{buzzi2019user} extended the cell-free approach to the 
case of a 
	user-centric massive MIMO approach and proposed power allocation strategies 
	aimed at either sum-rate maximization or minimum-rate maximization.
	\cite{zhang2019cell} provided a extensive survey of cell-free massive MIMO 
	systems and discussed the benefits of cell-free massive MIMO systems 
	including energy and cost efficiency.
\cite{ammar2021user} investigated the user-centric cell-free massive MIMO with 
distributed units to serve users. In such systems, a specific cluster of RAUs 
is 
served for each user and the user-centric scheme reduces edging effect, thus 
help to inprove the coverage and performance for users across the whole 
network.
\cite{wang2021live} proposed a cloud-based cell-free distributed massive MIMO 
system meet 5G NR requirements.
\cite{papazafeiropoulos2020performance} revealed the cell-free massive MIMO 
systems outmatch small-cells design for both coverage and rate because it takes 
advantage of both network MIMO and classical massive MIMO 
systems.} 
    Uplink SE in cell-free massive 
	MIMO systems is analyzed in \cite{ liu2019spectral}. The spacial correlated 
	propagation in cell-free massive MIMO with short-term power constrains is 
	studied in \cite{femenias2020short }.  \cite{zhang2019closed} analyzed the 
	uplink average ergodic capacity and gave a closed-form approximation in 
	distributed massive MIMO. {\cite{hu2017energy} investigated 
	an energy efficiency resource allocation scheme in downlink transmission to 
	maximize system energy efficiency considered power consumption including 
	transmitting power, calculation power and circuit power. 
	\cite{mosleh2019downlink} 
	proposed an iterative algorithm for	maximizing the minimum achievable rate 
	of downlink transmission among the UEs in cell-free massive 
	MIMO. \cite{amin2018quantized} proposesd two power allocation algorithms 
	based on quantizing the downlink transmitted powers of APs for reducing 
	complexity with Conjugate Beamforming (CB) and Zero-Forcing (ZF) 
	precoding.
	\cite{riera2019trade} evaluated different trade-offs between precoding 
	strategies, power allocation techniques and pilot allocation strategies 
	affecting the performance in cell-free massive MIMO networks. 
	\cite{izadi2020power} proposed an method for dividing the pilot 
	power and data power in the downlink transmission to maximize the minimum 
	achievable rate of the users in a cell-free massive MIMO system.
\cite{chakraborty2020efficient} proposed a new FP-based algorithm for spectral 
efficiency power allocation to solve several convex problems with closed-form 
solutions.\cite{chen2019dynamic} studied the power control and
allocation in uplink MIMO systems with different data traffic and developed a 
dynamic scheduling algorithm (DSA) with Lyapunov optimization techniques to 
optimize the long-term user throughput.
\cite{palhares2021robust} developed an robust minimum mean-square error(RMMSE) 
precoder iteratively to alleviate interference with imperfect channel state 
information (CSI) and an optimal and uniform power allocation schemes based on 
SINR.}
	Additionally, some works combined machine learning with distributed massive 
	MIMO and achieved a considerable improvement\cite{ding2021machine }.
{	These current works are mainly focus on  the performance 
analysis and 
	resource allocation of unicast in cell-free massive MIMO. }

Traditional unicast requires a large number of channels to transmit data when
the number of users is huge, which is inefficient and wasteful.
To improve the transmission efficiency, the theory of multicast
transmission called physical layer multicasting was first proposed in
\cite{sidiropoulos2006transmit}. Multicast is an efficient technique designed
for wireless communication to meet the demands of high data rate and low
latency. In multicast, the server can send a single data stream to the users
who need the same data and this scheme greatly reduces redundant data.
This idea was extended to multigroup multicast in \cite{karipidis2008quality}.
However, \cite{sidiropoulos2006transmit} and \cite{karipidis2008quality} both
simply assumed the CSI was known at the transmitter.

A multicast large-scale antenna system (also called massive MIMO system) was
considered in \cite{yang2013multicast} and the channel estimation for multicast
group was accomplished by a common pilot sequence for all the users who needed
the same data. In traditional unicast, the number of channels 
larger than the number of users need to be allocated for pilot transmission to 
ensure orthogonality, however,this new method allows more user
terminals to be supported with lower pilot cost. To reduce the computational
complexity, \cite{sadeghi2017reducing} proposed that by using the large number
of antennas, the intergroup interference could be cancelled.
Several optimization problems were investigated in previous studies.
An optimization problem of multigroup multicast under the per-antenna power
constraint was studied in \cite{christopoulos2015multicast}.
An optimal multicast beamforming structure focusing the max-min fairness (MMF)
problem was investigated in \cite{dong2020multi}.
A joint user scheduling and precoding for multigroup multicast with perfect
channel state information was studied in \cite{bandi2020joint} through a
convex-concave algorithm.
Joint unicast and multicast transmission in large-scale MIMO was studied
in \cite{sadeghi2018joint}, but only the centralized large-scale MIMO system.
These works all studied multicast in traditional centralized massive MIMO
systems.

Recently, multigroup multicast was extended to cell-free massive MIMO.
Conjugate beamforming was used in
\cite{doan2017performance,zhang2019max,tan2020energy} to analyze the
performance of multicast. The security of multigroup multicasting transmission
was studied in \cite{zhang2019secrecy}.
They all investigated the performance of multigroup multicast in cell-free
massive MIMO, however, to the best of our knowledge, the closed-form expression
of SE or signal-to-interference-plus-noise ratio (SINR) with different
precoding schemes under multicast transmission has not been analyzed in
cell-free massive MIMO scenarios yet.

This paper investigates the SE of joint unicast and multigroup multicast
transmission in cell-free distributed massive MIMO systems with different
precoding schemes. The main contributions of this paper are listed as follows:

\begin{enumerate}
 \setlength{\itemsep}{0pt}
 \setlength{\parsep}{0pt}
 \setlength{\parskip}{0pt}
\item We propose a joint unicast and multigroup multicast transmission scheme in
cell-free distributed massive MIMO systems. We derive the closed-form
expressions for
estimated CSI of unicast and multigroup multicast users in
cell-free distributed massive MIMO systems.
\item With the estimated CSI, the closed-form expressions of the SE of unicast
users and multigroup multicast users with  maximum ratio
transmission (MRT), zero-forcing (ZF) and minimum-mean-square error (MMSE)
precoding schemes in cell-free distributed massive MIMO are derived.
\item We fomulate a multi-objective optimization problem
(MOOP) between two conflicting objectives of maximizing the average SE of
unicast users and the MMF problem of multicast for joint unicast and
multigroup multicast transmission schemes in
cell-free distributed massive MIMO systems. The trade-off
regions between these two conflicting problems is obtained by
solving the MOOP with non-dominated sorting genetic algorithm II
(NSGA-\uppercase\expandafter{\romannumeral2}).
\item We convert the MOOP to a deep learning (DL) problem of
maximizing the sum of average achievable rate of unicast users
and minimum achievable rate of multicast users.  We propose an unsupervised
learning method to maximize the objective function by training the loss
function to the lowest. A deep neural network (DNN) is
designed to learn the nonlinear mapping between the input (channel large-scale
fading vector) and the output (power allocation scheme).
\item The accuracy of the derived closed-form expressions and the effectiveness
of the joint unicast and multigroup multicast transmission scheme in cell-free
distributed massive MIMO systems are verified. Insightful conclusions are drawn
from the SE analysis under various system parameters and the trade-off between
the considered two conflicting optimization objectives.
\end{enumerate}

The rest of this paper is organized as follows: System model consisting of
system configuration, channel model and channel estimation is presented in
Section \uppercase\expandafter{\romannumeral2}. Downlink data transmission
including downlink channel precoding is presented in Section
\uppercase\expandafter{\romannumeral3}. SE with three precoding schemes is
analyzed in Section \uppercase\expandafter{\romannumeral4}.
The efficient power allocation scheme with NSGA-II is investigated in
Section \uppercase\expandafter{\romannumeral5}.
	The power allocation scheme based on unsupervised DL is proposed in Section
	\uppercase\expandafter{\romannumeral6}.
Numerical result and discussions are presented in Section
\uppercase\expandafter{\romannumeral7}.  Finally, some conclusions are drawn in
Section \uppercase\expandafter{\romannumeral8}.

{\emph{Notation}}: Boldface letters in lower (upper) case denote vectors
(matrices). ${\mathbf{I}}_N$ means an $N$-dimensional identity matrix.
$(\cdot)^{\mathrm{H}}$ and $(\cdot)^{\mathrm{T}}$ represent the Hermitian
transpose and transpose, respectively. $|\cdot|$ and $\|\cdot\|$ represent the
absolute value and spectral norm, respectively. $\mathbb{E}[\cdot]$ denote
expectation. {$\otimes$ denotes the Kronecker product.} A
circularly symmetric complex Gaussian random variable $x$ with mean zero and
variance $\sigma^2$ is denoted as $x\sim\mathcal{CN}(0,\sigma^2)$.

\section{System Model}
We consider a cell-free distributed massive MIMO system that consists of a CPU,
several RAUs equipped with numerous antennas and a large number of unicast and
multicast user terminals. The CPU is used to design the precoding and process
the received signals and the RAUs collaborate with each other to jointly receive
and send signals.

\begin{figure}[t]
  \centering
  \includegraphics[width=8.5cm]{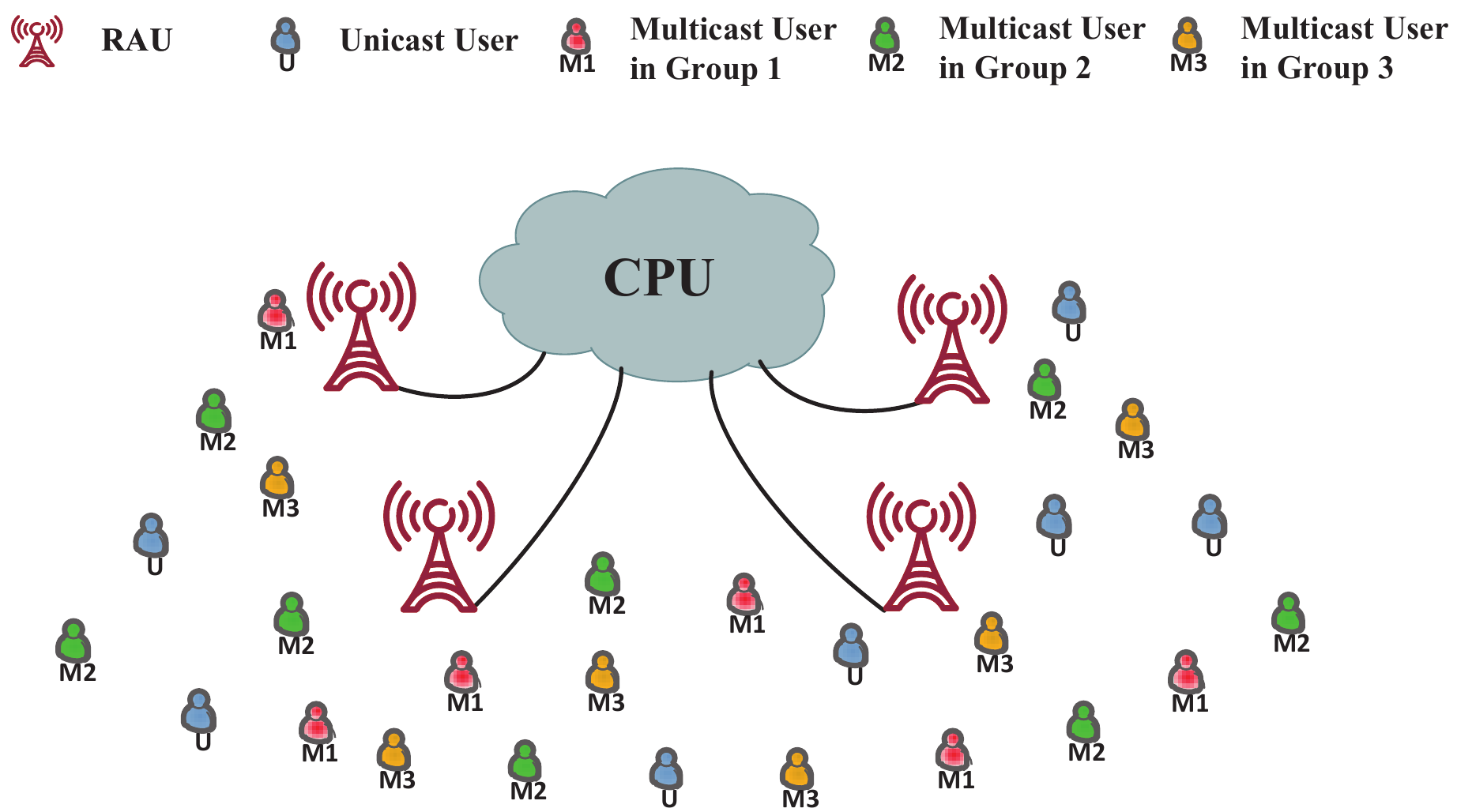}
  \caption{{Joint unicast and multicast in cell-free distributed
  massive MIMO systems.}\label{fig1}}
\end{figure}

\subsection{System Configuration}

We assume $N$ RAUs with $L$ antennas on each RAU, $U$ unicast users and $M$
multicast groups, among which the $m$-th multicast group contains
$K_{\mathrm{m}}$ multicast users, and all users and RAUs are randomly
distributed in the cell-free scenario. Fig. \ref{fig1} presents such a system.
During the joint unicast and multigroup multicast transmission, there are also
downlink interferences affect the performance of the system.

\subsection{Channel Model}
Considering frequency-flat fading channel, the channel vector between all RAUs
and the $u$-th unicast user or the $k$-th multicast user in $m$-th multicast
group can be described as
\begin{align}\label{cm}
\mathbf{c}_{u}&={\mathbf{\beta}_{u}^{1/2}\mathbf{h}_{u}}
{{\in {\mathbb{C}}^{NL \times 1}}},\\
\mathbf{t}_{m,k}&={\mathbf{\eta}_{m,k}^{1/2}\mathbf{h}_{m,k}}
{{\in {\mathbb{C}}^{NL \times 1}}},
\end{align}
where ${{\mathbf{\beta }}_{u}}=\mathbb{E}\left[
\mathbf{c}_{u}^{{}}\mathbf{c}_{u}^{\mathrm{H}} \right]=\text{diag}\left( \left[
{{\beta }_{1,u}},...,{{\beta }_{N,u}} \right] \right)\otimes
{{\mathbf{I}}_{L}}$ and ${{\mathbf{\eta }}_{m,k}}=\mathbb{E}\left[
\mathbf{t}_{m,k}^{{}}\mathbf{t}_{m,k}^{\mathrm{H}} \right]=\text{diag}\left(
\left[{ {{\eta }_{1,m,k}},...,} {{{\eta
}_{N,m,k}} }\right]
\right)\otimes{{\mathbf{I}}_{L}}$
refer to the channel covariance matrix,
${{\mathbf{\beta }}_{n,u}}$ and  ${{\eta }_{n,m,k}}$ represent the large-scale
fading between the $n$-th RAU and the $u$-th unicast user or the $k$-th
multicast user in the $m$-th multicast group,
${{\mathbf{h}}_{u}}\sim\mathcal{C}\mathcal{N}(0,{{\mathbf{I}}_{NL}})$ and
${{\mathbf{h}}_{m,k}}\sim\mathcal{C}\mathcal{N}(0,{{\mathbf{I}}_{NL}})$ are the
small-scale fast fading. It is assumed that the channels of 
different users are uncorrelated.

\subsection{Channel Estimation}
The system adopts time division duplex (TDD) mode.
Due to the reciprocity between uplink and downlink channels, the CSI obtained
by uplink channel estimation can be used for the precoding of
downlink data transmission. The system uses
uplink pilot sequences sent by users for channel estimation. Since all the user
terminals within a multicast group need the same data, they are combined to
share one pilot signal, thus only $M+U$ pilot signals need to be transmitted,
which greatly alleviates the pilot tension caused by the large number of
multicast users.

The uplink received pilot signal at CPU can be written as
\begin{align}\label{zrs}
\nonumber\mathbf{Y} = &\sum\limits_{u=1}^{U}{\sqrt{\tau p_{\mathrm{ul},u}}
	{{\mathbf{c}}_{u}}\mathbf{\phi}_{\mathrm{pu},u}^{\mathrm{H}}}\\ 
&+\sum\limits_{m=1}^{M}
{\sum\limits_{k=1}^{{{K}_{m}}}
	{\sqrt{\tau
	 {q}_{\mathrm{ul},m,k}}{{\mathbf{t}}_{m,k}}\mathbf{\phi}_{\mathrm{pm},m}
 ^{\mathrm{H}}}}
+\mathbf{N},
\end{align}
where $p_{\mathrm{ul},u}$ and ${q}_{\mathrm{ul},m,k}$
denote the uplink transmitted power of the $u$-th unicast user and the  $k$-th
multicast user in the $m$-th multicast group. 
${{\mathbf{\phi}}_{\mathrm{pu},u}}$ and
${{\mathbf{\phi}}_{\mathrm{pm},m}}$ represent the orthogonal pilot sequences of 
the
$u$-th unicast user and the $m$-th multicast group with length $\tau$. The
length of pilot sequences $\tau$ need to satisfy $\left( M+U \right)\le \tau
\le T$. To achieve lower pilot cost, $\tau$ can be chosen as $\tau = M+U $.
$\mathbf{N}$ is a  $NL\times \tau$ dimension complex additive white Gaussian
noise (AWGN).

The linear MMSE estimation\cite{bjornson2015massive} of ${\mathbf{{f}}}_{u}$
can be given by
\begin{align}\label{fkjc}
\nonumber	{{\mathbf{\hat{c}}}_{u}}=&\frac{\sqrt{\tau
			 p_{\mathrm{ul},u}}{{\mathbf{\beta
			}}_{u}}}{\tau
		 p_{\mathrm{ul},u}{{\mathbf{\beta }}_{u}}+\sigma
		_{\mathrm{ul}}^{2}{{\mathbf{I}}_{NL}}}\left( \sqrt{\tau
		 p_{\mathrm{ul},u}}{{\mathbf{c}}_{u}}+{{\mathbf{n}}_{u}}
		\right)\\
		=&\mathbf{\lambda
	}_{u}^{{1}/{2}\;}{{\mathbf{\hat{h}}}_{u}},
\end{align}
where   ${{\mathbf{\lambda
}}_{u}}=\text{diag}([{{\lambda
}_{1,u}},{{\lambda }_{2,u}},...,{{\lambda }_{N,u}}])
\otimes{\mathbf{I}_{L}}$
is the equivalent large-scale fading, ${{\lambda
}_{\mathrm{n,u}}}=\frac{\tau
 p_{\mathrm{ul},u}{{\beta
		}_{n,u}}^{2}}{\tau p_{\mathrm{ul},u}{{\beta }_{n,u}}+\sigma
	_{\mathrm{ul}}^{2}}$ and
${{\mathbf{\hat{h}}}_{u}}\sim\mathcal{C}\mathcal{N}(0,{{\mathbf{I}}_{NL}})$  is
the equivalent small-scale fast fading.

Similarly, the linear MMSE estimation of ${\mathbf{t}}_{m,k}$ can be given
by\begin{align}\label{ggkqc}
	{{\mathbf{\hat{t}}}_{m,k}}
	=\mathbf{\xi
	}_{m,k}^{{1}/{2}\;}{{\mathbf{\hat{h}}}_{m,k}},
\end{align}where
${{\mathbf{\xi
	}}_{m,k}}=\text{diag}\left( \left[ {{\xi }_{1,m,k}},...,{{\xi }_{N,m,k}}
\right] \right)\otimes {{\mathbf{I}}_{L}}$ is the equivalent large-scale
fading, ${{\mathbf{\xi}}_{n,m,k}}=\frac{\tau {q}_{\mathrm{ul},m,k}\eta
	_{n,m,k}^{2}}{\sigma
	_{\mathrm{ul}}^{2}+\sum\nolimits_{j=1}^{{{K}_{m}}}{\tau
		{q}_{\mathrm{ul},m,j}{{\eta }_{n,m,j}}}}$ and
	 ${{\mathbf{\hat{h}}}_{m,k}}\sim\mathcal{C}\mathcal{N}(0,{{\mathbf{I}}_{NL}})$.

By adopting co-pilot strategy, we can obtain the equivalent 
channel state information of the multicast group \cite{ yang2013multicast}. 
Compared with the case of fully orthogonal pilots, this method greatly saves 
pilot overhead. In addition, the same multicast group requires the same 
information to be transmitted, so the error due to pilot multiplexing is small.
To reduce pilot cost, we design a unified precoding for each multicast group. 
The co-pilot assignment strategy based channel
estimation of multicast groups $\mathbf{{t}}_{m}=	
\sum\nolimits_{j=1}^{{{K}_{m}}}{\sqrt{\tau
		{q}_{\mathrm{ul},m,j}}} {{\mathbf{t}}_{m,j}}$ is 
		adopted and the closed-form expression can be given by
\begin{align}\label{ggjc}	
\nonumber{{\mathbf{\hat{t}}}_{m}}=&\frac{\sum\nolimits_{j=1}^{{{K}_{m}}}{\tau
	 			{q}_{\mathrm{ul},m,j}{{\mathbf{\eta
				}}_{m,j}}}\left(
			\sum\nolimits_{j=1}^{{{K}_{m}}}{\sqrt{\tau
		 {q}_{\mathrm{ul},m,j}}{{\mathbf{t}}_{m,j}}}+{{\mathbf{n}}_{m}}
		\right)}{\sum\nolimits_{j=1}^{{{K}_{m}}}{\tau
			{q}_{\mathrm{ul},m,j}{{\mathbf{\eta }}_{m,j}}}+\sigma
		_{\mathrm{ul}}^{2}{{\mathbf{I}}_{NL}}}	\\
		=&\mathbf{\mu}_{m}^{{1}/{2}\;}{{\mathbf{\hat{h}}^*}_{m}},
\end{align}where
${{\mathbf{\mu }}_{m}}=\text{diag}\left( \left[
{{\mu}_{1,m}},{{\mu }_{2,m}},...,{{\mu }_{N,m}} \right] \right)\otimes
{{\mathbf{I}}_{L}}$  is the equivalent large-scale fading,
${{\mu}_{n,m}}=\frac{{{(\sum\nolimits_{j=1}^{{{K}_{m}}}{\tau
				{q}_{\mathrm{ul},m,j}{{\eta
					 }_{n,m,j}}})}^{2}}}{\sum\nolimits_{j=1}^{{{K}_{m}}}{\tau
		{q}_{\mathrm{ul},m,k}{{\eta }_{n,m,j}}}+\sigma _{\mathrm{ul}}^{2}}$ and
	 ${{\mathbf{\hat{h}}^*}_{m}}\sim\mathcal{C}\mathcal{N}(0,{{\mathbf{I}}_{NL}})$.
	
	\begin{remark}
	As seen from \eqref{ggkqc} and \eqref{ggjc},  the channel 
	estimation of multicast group $m$ can be given by  
	${{\mathbf{\hat{t}}}_{m}}=	
		\sum\nolimits_{j=1}^{{{K}_{m}}}{\sqrt{\tau
				{q}_{\mathrm{ul},m,j}}}{{\mathbf{\hat{t}}}_{m,j}}$  which is a 
				linear combination of the channel estimations of all users 
				within the $m$-th multicast group. 
			{Under the traditional orthogonal pilot 
			estimation scheme, each user in the multicast group need to be 
			assigned orthogonal pilots as unicast and the pilot overhead will 
			be extremely large.}
			As a result, the channel estimation	of multigroup
	multicast \eqref{ggjc} and unicast \eqref{fkjc} are completed respectively 
	to save pilot cost and
	increase accuracy simultaneously. 
	\end{remark}

\section{Downlink Data Transmission}
In this section, the downlink transmission of joint unicast and multigroup
multicast in cell-free distributed massive MIMO is studied. Since the 
transmission expressions of multicast and
unicast are relatively similar, only the expression of multigroup multicast in
joint
unicast and multigroup multicast transmission is given below.

The received signal of the $k$-th multicast user terminal in the $m$-th
multicast group can be presented as:
\begin{align}
\label{dlmcrs}
	\notag	{{z}_{m,k}} = &
		\mathbf{t}_{m,k}^{\mathrm{H}}(\mathbf{Wx}+\mathbf{Vs})+
		{{n}_{m,k}}\\
\nonumber	= &\mathbf{t}_{m,k}^{\mathrm{H}}{{\mathbf{w}}_{m}}{{x}_{m}}
		+\sum\limits_{r=1,r\ne
			 m}^{M}{\mathbf{t}_{m,k}^{\mathrm{H}}{{\mathbf{w}}_{r}}{{x}_{r}}}\\
	&	+\sum\limits_{u=1}^{U}{\mathbf{t}_{m,k}^{\mathrm{H}}
			{{\mathbf{v}}_{u}}{{s}_{u}}}+{{n}_{m,k}},
\end{align}
where  ${{n}_{m,k}}\sim
\mathcal{C}\mathcal{N}(0,\sigma
_{\mathrm{dl}}^{2})$ are complex AWGN vectors,
$\mathbf{x}={{\left[{{x}_{1}},...,{{x}_{M}} \right]}^{\text{T}}}$ and
$\mathbf{s}=\left[ {{s}_{1}},...,{{s}_{U}} \right]{}^{\mathrm{T}}$
are the data transmitted to multicast groups and unicast users.
$\mathbf{W}={{[{{\mathbf{w}}_{1}},{{\mathbf{w}}_{2}},...,
			{{\mathbf{w}}_{M}}]}^{\mathrm{T}}}$  and
			$\mathbf{V}={{[{{\mathbf{v}}_{1}},{{\mathbf{v}}_{2}},...,
		{{\mathbf{v}}_{U}}]}^{\mathrm{T}}}$ are the precoding
		matrixes of multicast groups and unicast users.
		
\subsection{Downlink Channel Precoding}
To facilitate the comparison of the performance under different precoding
schemes, we use the average power normalization criterion
\cite{mohammed2012single, yu2016alternating} to design the precoding vectors.
The precoding for each multicast group can be given by
\begin{align}
\label{wgpre}
	\mathbf{w}_{m}^{\mathrm{pre}}{=}\frac{\sqrt{q_{\mathrm{dl},m}}
			\mathbf{\hat{a}}{_{m}^{\mathrm{pre}}}}
{\mathbb{E}{{\left[
 \left(\mathbf{\hat{a}}{_{m}^{\mathrm{pre}}}\right)^{\mathrm{H}}
 \mathbf{\hat{a}}{_{m}^{\mathrm{pre}}}
	\right]}^{1/2}}},
\end{align}where
$\mathrm{pre}\in \left\{ \mathrm{MRT,ZF,MMSE} \right\}$,
$q_{\mathrm{dl},m}$ are the downlink power of multicast precoding vectors
and $\mathbb{E}\left[ {{\left\|\mathbf{w}_{m}^{\mathrm{pre}}
\right\|}^{2}} \right]={q}_{\mathrm{dl},m}$.
By replacing the corresponding items with unicast, the unicast precoding vector
$\mathbf{v}_{u}^{\mathrm{pre}}$ can be obtained.

The item ${\mathbb{E}{{\left[
\left(\mathbf{\hat{a}}{_{m}^{\mathrm{pre}}}\right)^{\mathrm{H}}\mathbf{\hat{a}}{_{m}^
{\mathrm{pre}}}\right]}}}$ in \eqref{wgpre} needs further
calculation.
For MRT precoding, we have $\mathbf{\hat{a}}{_{m}^{\mathrm{MRT}}}
={{\mathbf{\hat{t}}}_{m}}$,
\begin{align}
\mathbb{E}{{\left[
			 \left(\mathbf{\hat{a}}{_{m}^{\mathrm{MRT}}}\right)^{\mathrm{H}}
			 \mathbf{\hat{a}}{_{m}^{\mathrm{MRT}}}\right]}}
\nonumber	=&L\sum\limits_{n=1}^{N}{\frac{{{(\sum\nolimits_{j=1}^{{{K}_{m}}}
					{\tau q_{\mathrm{ul},m,j}{{\eta}_{n,m,j}}})}^{2}}}
				{{\sum\nolimits_{j=1}^{{{K}_{m}}}{\tau q_{\mathrm{ul},m,j}
			{{\eta}_{n,m,j}}}}+\sigma_{\mathrm{ul}}^{2}}}\\
			=&L\sum\limits_{n=1}^{N}{{{\mu}_{n,m}}}.
\end{align}

With the definition of ${\upsilon}_{m}=\sum_{n=1}^{N}{{{\mu}_{n,m}}}$, we have
\begin{align}
		\mathbf{w}_{m}^{\mathrm{MRT}}
			=\sqrt{\frac{q_{\mathrm{dl},m}}{L{{\upsilon
				}_{m}}}}{{\mathbf{\hat{t}}}_{m}}.
\end{align}

For ZF precoding, we have $\mathbf{\hat{a}}{_{m}^{\mathrm{ZF}}}
=\left[\mathbf{\hat{Q}(}{{\mathbf{\hat{Q}}}^{\mathrm{H}}}\mathbf{\hat{Q}}{{\mathbf{)}}^
	{\mathbf{-1}}}\right]_{U+m}$,
 where $\mathbf{\hat{Q}}=\left[ \mathbf{\hat{C},\hat{T}} \right]$,
 $\mathbf{\hat{C}}=\left[
{{{\mathbf{\hat{c}}}}_{1}},{{{\mathbf{\hat{c}}}}_{2}},...,{{{\mathbf{\hat{c}}}}_{U}}
 \right]$, $\mathbf{\hat{T}}=\left[
 {{{\mathbf{\hat{t}}}}_{1}},{{{\mathbf{\hat{t}}}}_{2}},...,{{{\mathbf{\hat{t}}}}
 	_{M}}\right]$, $\left[ \cdot \right]_{i}$ represents the $i$-th column of
 	the matrix,
 {
\begin{align}
\mathbb{E}{{\left[
		\left(\mathbf{\hat{a}}{_{m}^{\mathrm{ZF}}}\right)^{\mathrm{H}}
		\mathbf{\hat{a}}{_{m}^{\mathrm{ZF}}}
		\right]}}
\notag	=&\mathrm{tr} \left(\mathbf{\hat{Q}}^{\mathrm{H}} 
\left(	{{\left({\mathop{{{{\mathbf{\hat{Q}}}}^{\mathrm{H}}}
				\mathbf{\hat{Q}}}}
		\right)}^{-1}}\right)^{\mathrm{H}} \times 
{{\left({\mathop{{{{\mathbf{\hat{Q}}}}^{\mathrm{H}}}
				\mathbf{\hat{Q}}}}
		\right)}^{-1}} \mathbf{\hat{Q}}\right)
\\
\notag  =&{{((NL-M-U){{\upsilon}_{m}})}^{\text{-2}}} 
\mathrm{tr}\left(\mathbf{\hat{Q}}^{\mathrm{H}} \mathbf{\hat{Q}}\right)\\
=&{({(NL-M-U){{\upsilon }_{m}}})^{-1}},
\end{align}}
and the precoding vector can be given by
\begin{align}
		 \mathbf{w}_{m}^{\mathrm{ZF}}
&=\sqrt{A q_{\mathrm{dl},m}{{\upsilon}_{m}}}
	 \left[\mathbf{\hat{Q}(}{{{\mathbf{\hat{Q}}}}^{\mathrm{H}}}
	\mathbf{\hat{Q}}{{\mathbf{)}}^{\mathbf{-1}}}\right]_{U+m},
\end{align}
where $A=NL-M-U$.

For MMSE precoding, with the estimated vector
$\mathbf{\hat{a}}{_{m}^{\mathrm{MMSE}}}
=\left[\mathbf{\hat{Q}}{{\left({\mathop{{{{\mathbf{\hat{Q}}}}^{\mathrm{H}}}
	\mathbf{\hat{Q}}+\sigma_{\mathrm{dl}}^{2}}}\,\otimes
 {\mathbf{I}_{NL}}\right)}^{-1}}\right]_{U+m}$,
the pending item can be given by
\begin{align}
\notag&\mathbb{E}{{\left[
		\left(\mathbf{\hat{a}}{_{m}^{\mathrm{MMSE}}}\right)^{\mathrm{H}}
		\mathbf{\hat{a}}{_{m}^{\mathrm{MMSE}}}\right]}}\\
\notag	=&\mathrm{tr}\Big(
\begin{aligned}
&\mathbf{\hat{Q}}^{\mathrm{H}} 
\left(	{{\left({\mathop{{{{\mathbf{\hat{Q}}}}^{\mathrm{H}}}
				\mathbf{\hat{Q}}+\sigma_{\mathrm{dl}}^{2}}}\,\otimes
		{\mathbf{I}_{NL}}\right)}^{-1}}\right)^{\mathrm{H}} \\
&\times {{\left({\mathop{{{{\mathbf{\hat{Q}}}}^{\mathrm{H}}}
				\mathbf{\hat{Q}}+\sigma_{\mathrm{dl}}^{2}}}\,\otimes
		{\mathbf{I}_{NL}}\right)}^{-1}} \mathbf{\hat{Q}}
\end{aligned}\Big)  \\
\notag  =&{{(A{{\upsilon}_{m}}+\sigma_{\mathrm{dl}}^{2})}^{\text{-2}}} 
  \mathrm{tr}\left(\mathbf{\hat{Q}}^{\mathrm{H}} \mathbf{\hat{Q}}\right)\\
=&{{{(A{{\upsilon}_{m}}+\sigma_{\mathrm{dl}}^{2})}^{\text{-2}}}A{\upsilon}_{m}},
\end{align}

and the precoding vector can be given by
\begin{align}
\notag &\mathbf{w}_{m}^{\mathrm{MMSE}}\\
=&\sqrt{\frac{q_{\mathrm{dl},m}
		{{\left(A{{\upsilon}_{m}}+\sigma _{\text{dl}}^{2}\right)}^{2}}}
{A{{\upsilon}_{m}}}}\left[\mathbf{\hat{Q}}{{\left(
	{\mathop{{{{\mathbf{\hat{Q}}}}^{\mathrm{H}}}\mathbf{\hat{Q}}+\sigma_{\mathrm{dl}}^{2}}}\otimes
		 {\mathbf{I}_{NL}}\right)}^{-1}}\right]_{U+m}.
\end{align}

The corresponding precoding vectors of unicast $\mathbf{v}_{u}^{\mathrm{pre}} $
can be obtained by replacing
$q_{\mathrm{dl},m}$ and ${{\upsilon}_{m}}$ with the downlink power of unicast
precoding vectors $p_{\mathrm{dl},u}$ and the estimated covariance matrix of
unicast ${{\theta }_{u}}=\sum_{n=1}^{N}{\frac{\tau
p_{\mathrm{ul},u}{{\beta}_{n,u}^{2}}}{\tau p_{\mathrm{ul},u}{{\beta }_{n,u}}
+\sigma_{\mathrm{ul}}^{2}}}$ and $\mathbb{E}\left[ {{\left\|
\mathbf{v}_{u}^{\mathrm{pre}}\right\|}^{2}} \right]=p_{\mathrm{dl},u}$.

\begin{remark}
	Different from pure unicast or multicast, the joint unicast and multicast
	transmission system not only retains the individualization of unicast, but
	also promotes the efficiency of transmitting repeated information by
	multicast.
\end{remark}
\section{Spectral Efficiency Analysis}
In this section, we analyze the downlink achievable SE of user terminals with
different precoding schemes in joint unicast and multigroup multicast
transmission systems under
cell-free distributed massive MIMO scenarios. According to the standard
capacity bounding technique in
\cite{sadeghi2018joint,ngo2013energy,larsson2016joint},
the received signal from the RAUs to  the $k$-th unicast user and the $k$-th
multicast user in the $m$-th multicast group can be rewritten as
\begin{align}
\nonumber	{{y}_{k}}=&\mathbb{E}\left[
			\mathbf{c}_{k}^{\mathrm{H}}{{\mathbf{v}}_{k}}
			\right]{{s}_{k}}+\left(
			 \mathbf{c}_{k}^{\mathrm{H}}{{\mathbf{v}}_{k}}-\mathbb{E}\left[
			\mathbf{c}_{k}^{\mathrm{H}}{{\mathbf{v}}_{k}} \right]
			\right){{s}_{k}}\\
			&+\sum\limits_{u=1,u\ne
			k}^{U}{\mathbf{c}_{k}^{\mathrm{H}}
	{{\mathbf{v}}_{u}}{{s}_{u}}}+\sum\limits_{r=1}^{M}
	{\mathbf{c}_{k}^{\mathrm{H}}{{\mathbf{w}}_{r}}{{x}_{r}}}+{{n}_{k}},\\
\nonumber	{z}_{m,k}=&\mathbb{E}\left[
		\mathbf{t}_{m,k}^{\mathrm{H}}{{\mathbf{w}}_{m}}\right]{{x}_{m}}
		+\left(
		 \mathbf{t}_{m,k}^{\mathrm{H}}{{\mathbf{w}}_{m}}-\mathbb{E}\left[
		\mathbf{t}_{m,k}^{\mathrm{H}}{{\mathbf{w}}_{m}} \right]
		\right){{x}_{m}} \\
		&+\sum\limits_{r=1,r\ne
		m}^{M}{\mathbf{t}_{m,k}^{\mathrm{H}}
	{{\mathbf{w}}_{r}}{{x}_{r}}}
+\sum\limits_{u=1}^{U}{\mathbf{t}_{m,k}^{\text{H}}
			{{\mathbf{v}}_{u}}{{s}_{u}}}+{{n}_{m,k}},
\end{align}
where $\mathbb{E}\left[
\mathbf{c}_{k}^{\mathrm{H}}{{\mathbf{v}}_{k}}
\right]{{s}_{k}}$, $\mathbb{E}\left[
\mathbf{t}_{m,k}^{\mathrm{H}}{{\mathbf{w}}_{m}} \right]{{x}_{m}}$ can be
regarded as the signals needed and the others are interferences and noises.

In terms of \cite{li2017downlink}, the achievable SE can be
given by
\begin{align}	\label{sinr}
	& \text{SE}_{}^{\mathrm{pre}}=\left( 1-\frac{\tau }{T} \right)
	{{\log}_{2}}\left( 1+\text{SINR}_{}^{\mathrm{pre}} \right),
\end{align}
where the SINR of unicast and multicast can be represented as \eqref{sinrun}
and \eqref{sinrmu}.
\begin{figure*}[t]
\begin{align}\label{sinrun}
\text{SINR}_{k,un}^{\mathrm{pre}}&=\frac{{{\left|
	\mathbb{E}\left[
		 \mathbf{c}_{k}^{\mathrm{H}}\mathbf{v}_{k}^{\mathrm{pre}}
	\right]\right|}^{2}}}{\sigma _\mathrm{ dl}^{2}
-{{\left|\mathbb{E}\left[
		\mathbf{c}_{k}^{\mathrm{H}}\mathbf{v}_{k}^{\mathrm{pre}}
	\right]\right|}^{2}}
+\sum\limits_{u=1}^{U}{\mathbb{E}\left[{{\left|
		\mathbf{c}_{k}^{\mathrm{H}}\mathbf{v}_{u}^{\mathrm{pre}}
		\right|}^{2}}\right]}
+\sum\limits_{r=1}^{M}{\mathbb{E}\left[ {{\left|
		\mathbf{c}_{k}^{\mathrm{H}}\mathbf{w}_{r}^{\mathrm{pre}}
		 \right|}^{2}}\right]}} ,\\
\label{sinrmu}
\text{SINR}_{m,k,mu}^{\mathrm{pre}}
&=\frac{{{\left|\mathbb{E}\left[
		\mathbf{t}_{m,k}^{\text{H}}\mathbf{w}_{m}^{\text{pre}} \right]
		\right|}^{2}} }{\sigma_{\mathrm{dl}}^{2}-{{\left|
		\mathbb{E}\left[\mathbf{t}_{m,k}^{\text{H}}\mathbf{w}_{m}^{\text{pre}}
		 \right]\right|}^{2}}
	+\sum\limits_{u=1}^{U}{\mathbb{E}\left[
		{{\left|\mathbf{t}_{m,k}^{\text{H}}\mathbf{v}_{u}^{\text{pre}}
		\right|}^{2}}
		\right]}+\sum\limits_{r=1}^{M}{\mathbb{E}\left[
		{{\left| \mathbf{t}_{m,k}^{\text{H}}\mathbf{w}_{r}^{\text{pre}}
		\right|}^{2}}	\right]} },
\end{align}
	\begin{align}\label{closeun}
	\text{SINR}_{k,un}^{\text{pre}}
	&	=\frac{Jp_{\text{dl},k}{{\theta}_{k}}}{\sigma _{\text{dl}}^{2}
		+\sum\limits_{u=1}^{U}\left[\frac{1}{NL}{p_{\text{dl},u}\sum\limits_{n=1}^{N}
			{{\beta}_{n,k}}}
		\right]+\sum\limits_{r=1}^{M}{\left[\frac{1}{NL}q_{\mathrm{dl},r}
			\sum\limits_{n=1}^{N}{{{\beta }_{n,k}}}\right]}},\\
	\label{closemu}
	\text{SINR}_{m,k,mu}^{\mathrm{pre}}
	&	=\frac{J q_{\mathrm{dl},m}{{\zeta}_{m,k}}}{\sigma_{\mathrm{dl}}^{2}
		+\sum\limits_{u=1}^{U}{\left[\frac{1}{NL}
			p_{\mathrm{dl},u}\sum\limits_{n=1}^{N}{{{\eta}_{n,m,k}}}\right]}
		+\sum\limits_{r=1}^{M}{\left[\frac{1}{NL}q_{\mathrm{dl},r}\sum\limits_{n=1}^{N}
			{\eta _{n,m,k}}\right]}},
\end{align}
	\rule[-0pt]{18.2cm}{0.05em}
\end{figure*}

\begin{theorem}
	Each item can be solved with different precoding schemes obtained in
	the previous section and the closed-form expressions of achievable
	SINR with different precoding can be given by \eqref{closeun} and
	\eqref{closemu},
where
${{\zeta }_{m,k}} = \sum_{n=1}^{N}{\frac{\tau
q_{\mathrm{ul},m,k}\eta _{n,m,k}^{2}}{\sigma_{\mathrm{ul}}^{2}
+\sum\nolimits_{j=1}^{{{K}_{m}}}{\tau
q_{\mathrm{ul},m,j}{{\eta}_{n,m,j}}}}} $,

\begin{align}
	\notag J=\left\{ \begin{aligned}
		&
		L,\quad\quad\quad\quad\quad\quad\quad\quad\quad\quad\,\,\,\,\text{for
			MRT}, \\
	& NL-M-U,\quad\quad\quad\quad\quad\,\,\,\text{for  ZF}, \\
	& {{\left( NL \right)}^{2}}/\left( NL-M-U \right),\quad\text{for  MMSE}.
	\end{aligned} \right.
\end{align}
{\emph{Proof: Please refer to Appendix \ref{ap}.}}
\end{theorem}

	\begin{remark}
	In cell free distributed massive MIMO scenarios, due to the operation
		between RAUs,
		the SE of unicast and multicast can be greatly promoted and since the 
		closed-form expressions we derived only rely on large-scale fading 
		parameters which change much slower than small-scale parameters, this 
		is much highly practical. The complex
		formulas in distributed scenario can be transformed into simpler forms 
		in centralized scenarios by setting the number of RAUs 
		$N=1$ while the formulas for centralized scenarios 
		cannot be easily applied to distributed scenarios.
	\end{remark}

\section{Joint Spectral Efficiency Optimization}
{The derived closed-form expression of achievable SE (also 
SINR) shows that SE
can be influenced by downlink transmitted power, uplink pilot power and length 
of the pilot sequence $\tau$ .} To achieve higher SE, the
allocation of these resources needs to be considered, subject to various
constraints. For multicast, a problem worth studying is the MMF, in which the
minimum SINR or SE of the system needs to be maximized with
a limited transmitted power. However, due to the interference
between unicast and multicast, the SE of multicast and unicast users are
conflicting with each other, i.e. with the increase of the downlink power of
multicast, the SE of unicast users will decrease and vice versa, which means
that multicast and unicast users can not achieve the highest SE
simultaneously.
Therefore, a multi-objective optimization problem(MOOP) is formulated. One
objective
is the average achievable SE of unicast users and the other one is the MMF
problem for multicast users. Finally, we solve this MOOP problem with
NSGA-\uppercase\expandafter{\romannumeral2}.
\subsection{Problem Formulation}
For multicast, the MMF problem can be described as
\begin{align} \label{mmf}
	\underset{\mathcal P,\tau}{\mathop
	 \text{max}}\quad {f1}=& \min \,\,\,
	\text{SE}_{m,k,mu}^{\text{pre}} \\
	{\rm{s}}{\rm{.t}}{\rm{.}} \quad
\nonumber	C1: &\text{SE}_{m,k,mu}^{\text{pre}}\ge
	\text{SE}_{m,k,mu}^{\text{pre,min}},\,\forall m\in \mathcal M,\forall k\in
	{\mathcal{K}_{m}},\\
&	\text{SE}_{k,un}^{\text{pre}}\ge
	\text{SE}_{k,un}^{\text{pre,min}},\forall u\in \mathcal U, \\
\nonumber C2: &{{p}_{\mathrm{ul},k}}\le {{P}_{\mathrm{ul},un}},
\forall u\in\mathcal U,\label{C2}\,\,\,\\
 &{{q}_{\mathrm{ul},m,k}}\le {{P}_{\mathrm{ul},mu}},
\forall m\in \mathcal M,\forall k\in {\mathcal{K}_{m}}, 	\\
 \label{C3}C3:&\sum_{k=1}^{U}{{{p}_{\mathrm{dl},k}}}\le
{{P}_{\mathrm{dl},un}},\,
\sum_{m\text{=}1}^{M}{{{q}_{\mathrm{dl},m}}}\le
{{P}_{\mathrm{dl},mu}},\\
&\sum_{k=1}^{U}{{{p}_{\mathrm{dl},k}}}+\sum_{m\text{=}1}^{M}
{{{q}_{\mathrm{dl},m}}} \le {{P}_{\mathrm{dl}}}, \\
\label{C4}C4:&{{p}_{\mathrm{ul},k}}\ge 0,{{q}_{\mathrm{ul},m,k}}\ge
0,{{p}_{\mathrm{dl},k}}\ge
	0,{{q}_{\mathrm{dl},m}}\ge 0, \\
C5:&\tau \in \left\{ M+U,...,T \right\},\label{C5}
\end{align}
where $\mathcal P=\left\{
{{\text{p}}_{\text{ul}}}\text{,}{{\text{q}}_{\text{ul}}}\text{,}{{\text{p}}_{\text{dl}}}\text{,}{{\text{q}}_{\text{dl}}}
 \right\}$, ${{\text{p}}_{\text{ul}}}=\left[ {{p}_{\mathrm{ul},1}},\cdots
 ,{{p}_{\mathrm{ul},U}}\right]$,
 ${{\text{p}}_{\text{dl}}}=
 \left[ {{p}_{\mathrm{dl},1}},\cdots,{{p}_{\mathrm{dl},U}}\right]$,
 ${{\text{q}}_{\text{dl}}}=\left[
 {{q}_{\mathrm{dl},1}},\cdots,{{q}_{\mathrm{dl},M}}\right]$,
 ${{\text{q}}_{\text{ul}}}=\left[
 {{q}_{\mathrm{ul},1,1}},\cdots,{{q}_{\mathrm{ul},1,{{K}_{1}}}};\cdots;
 {{q}_{\mathrm{ul},M,1}},\cdots,{{q}_{\mathrm{ul},M,{{K}_{M}}}}\right]$.
{$C1$ are the ergodic downlink QoS constraints where
$\text{SE}_{m,k,mu}^{\text{pre,min}}>0$ and
$\text{SE}_{k,un}^{\text{pre,min}}>0$ are the required minimum SE of multicast
and unicast users.} $C2$ are the uplink power constraints where
${{P}_{\mathrm{ul},un}}$ and ${{P}_{\mathrm{ul},mu}}$ are the upper limits of
the uplink transmitted power per user of unicast and multicast. $C3$ are the
total downlink power constraints of all RAUs where ${{P}_{\mathrm{dl},un}}$ and
${{P}_{\mathrm{dl},mu}}$ are the upper limits of total downlink transmitted
power of unicast and multicast and ${{P}_{\mathrm{dl}}}$ is the total downlink
power threshold.
According to the discussion about the pilot sequence
length in Section II, we can give the optimal value of $C5$ 
is $\tau^*=M+U$.
Due to the non-convexity of $\text{SE}_{m,k,mu}^{\text{pre}}$, it is very
difficult to find the maximum value.

For unicast, the average SE optimization problem can be formulated as
\begin{align} \label{ase}
\underset{\mathcal P,\tau}{\mathop
	 \text{max}}\quad\,&{f2}=\sum\limits_{u=1}^{U}{\text{SE}_{u,un}^{\text{pre}}}/U,
	\\ 		{\rm{s}}{\rm{.t}}{\rm{.}}\quad \,\,&C1-C5.
\end{align}

The MMF problem of multicast $\mathop {f1}$ and the maximization of average
achievable SE of unicast $\mathop {f2}$ are conflicting.
Thus we propose a MOOP to investigate the trade-off between these two problems
\begin{align}\label{moop}
	\underset{\mathcal P,\tau}{\mathop \text{max}}\,\,\quad& \mathbf{F} {=
	}{{\left[ f1,f2 \right]}^{\text{T}}},\\
 	{\rm{s}}{\rm{.t}}{\rm{.}}\quad \,\,&C1-C5,
\end{align}
where $\mathbf{F}$ is the optimal vector containing the objective functions $f1$
and $f2$. The
MOOP aims to maximize the minimum SE of multicast and the average SE of unicast
simultaneously.
\subsection{Solutions by Multi-objective Genetic Algorithm}
Most optimization methods transform a
MOOP into multiple single-objective optimizations by giving one Pareto-optimal
solution at a run. However, these kinds of methods may obtain different
solutions at different runs and not give a joint optimal solution. 
$f1$ and $f2$ are both non-convex problems due to the 
non-convexity of $	\text{SE}_{m,k,mu}^{\text{pre}} $ and 
$\text{SE}_{u,un}^{\text{pre}}$ . 
In order to solve the two objectives simultaneously in a single 
simulation run, the multi-objective evolutionary algorithms (MOEAs) is utilized 
to obtain the Pareto boundary of \eqref{moop}. Therefore, we adopt MOEAs to 
obtain the simultaneous Pareto-optimal solutions 
in this paper. Specifically, NSGA-\uppercase\expandafter{\romannumeral2} 
\cite{deb2002fast} which has been proved to be reliable and effective in 
\cite{WS2018optimization} is utilized to solve the proposed MOOP.

NSGA-II is an elitist nondominated sorting genetic algorithm with a lower
complexity of $O(XY^2)$, where $X$ is the number of objective functions and $Y$
is the population size. According to NSGA-II, firstly, an
initial population with a size of $Y$ is randomly generated. In our case, the
uplink pilot power, the downlink transmitted power and the coherence time are
selected under the constraints given in \eqref{C3} - \eqref{C5}.
Secondly, the two objective functions are evaluated based on the current
population by calculating the minimum SE of multicast users and the average SE
of unicast users as \eqref{mmf} and \eqref{ase}. Thirdly, we rank the
population based on fast non-dominated sorting and crowdedness.
Then a new parent population is selected from the
currently ranked population by tournament selection and an offspring population	
is generated by selection, mutation, and crossing. Finally, the offspring
population and initial population are combined together and ranked to a new
initial population. The cycle repeats until the optimization criteria are met 	
and the Pareto boundary of the MOOP can be found. The power 
allocation algorithm based on NSGA-II is shown in Algorithm 1.

{
	\begin{algorithm}[h]
		\caption{Power Allocation Algorithm Based on 
		NSGA-\uppercase\expandafter{\romannumeral2}}
		\label{NSGA allocation algorithm}
		\begin{algorithmic}
			\STATE{\textbf{Initialization:}}
			\begin{itemize}
				\item initialize system parameters:  the uplink transmitted 
				power of unicast and mulicast ${{\text{p}}_{\text{ul}}}$, 
				${{\text{q}}_{\text{ul}}}$, the downlink transmitted power of 
				unicast and mulicast ${{\text{p}}_{\text{dl}}}$, 
				${{\text{q}}_{\text{dl}}}$, 
				the size of population $Y$ and so on
				\item  generate the population randomly under the constraints 
				given in \eqref{C3} - \eqref{C5} 
			\end{itemize}
			\STATE{\textbf{Evaluation:}}
			\begin{itemize}
				\item evaluate the two objective functions: the minimum SE of 
				multicast users and the average SE of unicast users as 
				\eqref{mmf} and \eqref{ase}
				\item select an excellent parent population to generate 
				offspring population by selection, mutation, and crossing
			\end{itemize}
			\STATE{\textbf{Sort:}}
			\begin{itemize}
				\item rank each individual based on non-dominating 
				sorting and crowding distance
			\end{itemize}
		\STATE{\textbf{Repeat:}}
		\begin{itemize}
			\item select an excellent parent population to generate offspring 
			population by selection, mutation, and crossing
		\end{itemize}
		\begin{itemize}
			\item rank the $2Y$ combined offspring population and the old 
			population to generate a new population
		\end{itemize}
	        \STATE{\textbf{Until}} the stop criterion of optimization 
			\STATE{\textbf{Return}} a series of Pareto optimal solutions of 
			(\ref{moop})
		\end{algorithmic}
	\end{algorithm}
}

\section{Unsupervised Deep Learning Based Power Allocation}
Given the proposed MOOP in \eqref{moop}, to further reduce the computational
complexity, we introduce an unsupervised DL method to optimize the achievable
rate of multicast and unicast users and compare it with NSGA-II algorithm.

\subsection{Network Design}
The traditional DL approach is to decrease the loss function
continuously and minimize the loss function \cite{lecun2015deep,8283585}. To
meet the
demands of maximizing
the average SE of unicast and the minimum SE of multicast users, we take the
negative SE as the loss function, so while the loss function is decreasing, the
SE is increasing. The loss function can be given by
\begin{align}
\nonumber	\underset{\mathcal P}{ loss} =&\mathbb{E}\left[ 
	los{{s}_{ase}}+los{{s}_{mmf}} \right]\\
 =&\mathbb{E}\left[
-\sum\limits_{u=1}^{U}{\text{SE}_{k,un}^{\text{pre}}}/U-\min \left(
\text{SE}_{m,k,mu}^{\text{pre}} \right) \right].
\end{align}

Through the optimization of the coherence time $T$ with NSGA-II, we can obtain
the approximately optimal value $T^*$ and the loss function can be given by
\begin{align}
\nonumber&	\underset{\mathcal P} {loss}\\ =&\mathbb{E}\Big[\left( 
1-\frac{M+U}{{{T}^{*}}}
\right)
\Big(
\begin{aligned}
&	-\sum_{u=1}^{U}{{{\log }_{2}}\left( 1+\text{SINR}_{k,un}^{\text{pre}}
		\right)}/U\\
&	-\min {{\log}_{2}}\left( 1+\text{SINR}_{m,k,mu}^{\text{pre}} \right)
\end{aligned}
\Big)
 \Big].
\end{align}

Since some of the optimization variables have been determined, the optimization
problem can be transformed into maximizing the achievable rate and loss
function
\begin{align}
	\underset{\mathcal P} {loss}^{'}=\mathbb{E}\Big[
	\begin{aligned}
&-\sum\limits_{u=1}^{U}{{{\log }_{2}}\left(
	1+\text{SINR}_{k,un}^{\text{pre}} \right)}/U \\
&-\min\left( {{\log
	}_{2}}\left( 1+\text{SINR}_{m,k,mu}^{\text{pre}} \right)\right)
	\end{aligned}
\Big].
\end{align}
\subsection{Network Structure}
In this part, we consider a fully connected deep neural network (DNN) to
maximize the achievable rate of unicast and multicast users. The input of DNN
are the channel large-scale fading of the system $\mathbf{\beta}_{u}$ and
$\mathbf{\eta}_{m,k}$. The output of DNN is the power allocation strategy
$\mathcal P$ to maximize the achievable SE, where the first $ 
U+M $ outputs
denote the optimal downlink transmitted power of unicast
and multicast and the following outputs denote the optimal uplink pilot power.
The DNN is trained to learn the nonlinear mapping between the large-scale
fading and the power allocation scheme.

  \begin{figure}
 	\centering
 	\includegraphics[width=8.5cm]{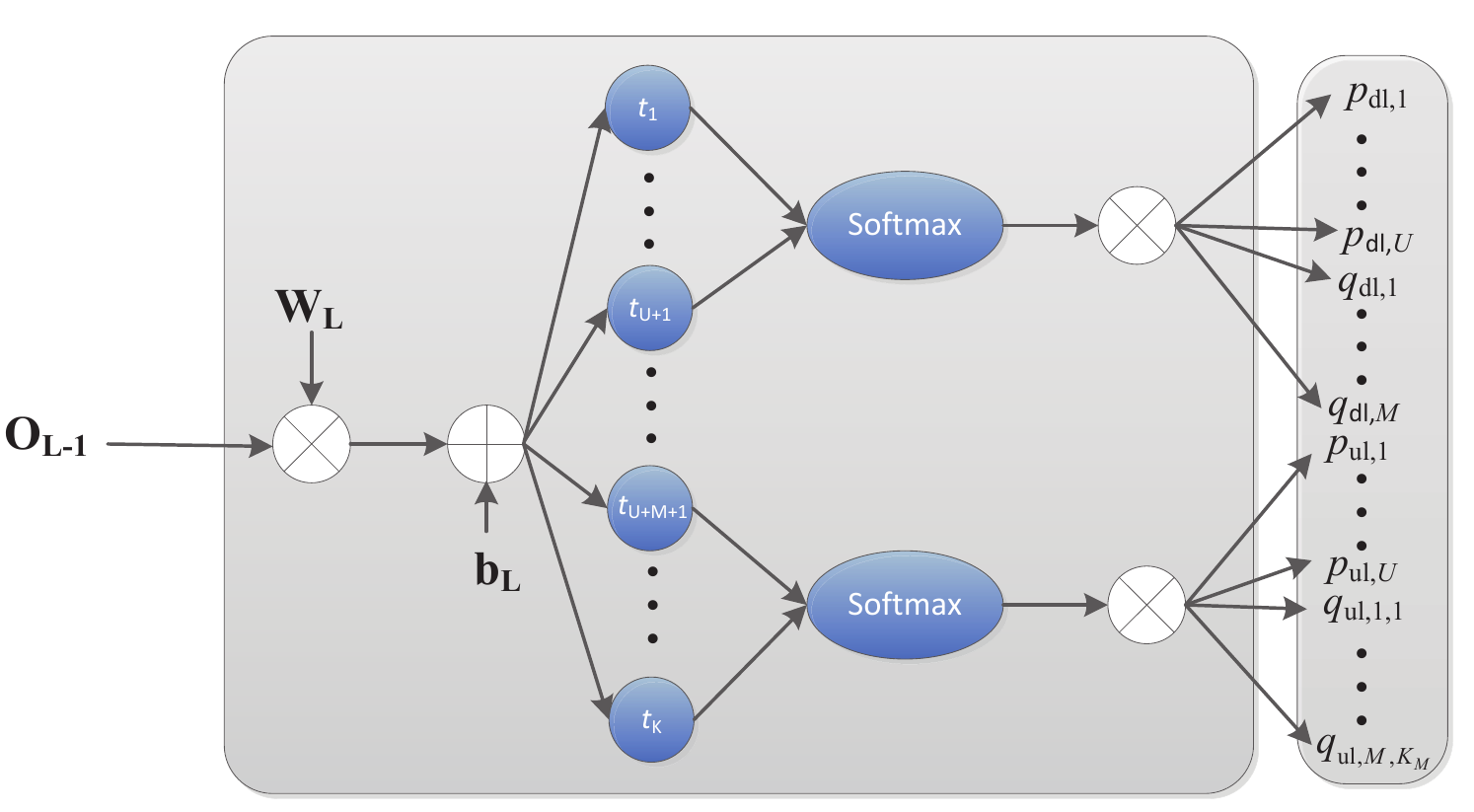}
 	\caption{{The detailed structure of the output layer of the unsupervised
 	learning based DNN.}\label{fig10}}
 \end{figure}
The detailed structure of the output layer is shown in Fig. \ref{fig10}, where
${{\mathbf{o}}_{L-1}}$ is the output of the hidden layer. ${{\mathbf{W}}_{L}}$
is the weight matrix and ${{\mathbf{b}}_{L}}$ is the bias vector of layer $L$,
which needs to be optimized in neural networks. Processed by a linear
transformation
$\mathbf{g}={{\mathbf{W}}_{L}}{{\mathbf{o}}_{L-1}}+{{\mathbf{b}}_{L}}$, a
temporary output $\mathbf{g}$ can be obtained. Then $\mathbf{g}=\left[
g_{1}^{{}},\cdots ,{{g}_{K}} \right]$ is divided into two groups. One is from
${{g}_{1}}$ to
${{g}_{M+U}}$ and the other is from ${{g}_{M+U+1}}$ to ${{g}_{K}}$ where
$K=2U+M+\sum_{m=1}^{M}{{{K}_{m}}}$. In order to achieve
the power constraints proposed in \eqref{C2} - \eqref{C4}, each group is first
submitted to a
Softmax function to achieve normalization and then multiplies the power
constraints of the group, which can be expressed as
\begin{align}
{{p}_{k}}=\left\{ \begin{aligned}
	& p_{\text{dl}}^{\text{tot}}\sigma \left( {{g}_{k}} \right),
	\quad 1\le k\le M+U, \\
	& p_{\text{ul}}^{\text{tot}}\sigma \left( {{g}_{k}} \right),
	\quad M+U+1\le 	k\le K, \\
\end{aligned} \right.
\end{align}
where \begin{align}
 \sigma \left( {{t}_{k}} \right)=\left\{ \begin{aligned}
	&
	\frac{{{e}^{{{t}_{k}}}}}{\sum\nolimits_{k=1}^{M+U}{{{e}^{{{t}_{k}}}}}},
   \quad \quad\,\, 1\le k\le M+U, \\
	&
	\frac{{{e}^{{{t}_{k}}}}}{\sum\nolimits_{k=M+U+1}^{K}{{{e}^{{{t}_{k}}}}}},
    \,\, M+U+1\le k\le K, \\
\end{aligned} \right.
\end{align}
and $p_{\text{dl}}^{\text{tot}}=\sum_{k=1}^{M+U}{{{p}_{k}}}$,
$p_{\text{ul}}^{\text{tot}}=\sum_{k=M+U{+}1}^{K}{{{p}_{k}}}$
are the total power constraints of uplink and downlink.

We adopt the Adaptive Moment Estimation (Adam) algorithm for optimization,
which is essentially an RMSprop algorithm with a momentum term. Adjusting the 
learning rate of each parameter, it can minimize the loss function
and maximize the optimization objective. The main advantage of Adam is that
after bias correction, each iteration of the learning rate has a certain range,
which makes the parameters relatively stable.

The complexity of training phase can be regarded as $O\left( 2\left( 
KN{{n}_{1}}+K\tau {{n}_{L-1}}+\sum\limits_{l=2}^{L-1}{{{n}_{l-1}}{{n}_{l}}} 
\right)st \right)$, where $s$ is the size of mini-batch, $t$ is the number 
of iterations and $n_l$ denotes the number of nodes of the $l$-th layer.
Through the training of the DNN with the large-scale fading coefficients, the 
considered DL problem can be solved. Therefore, the complexity of solving the 
DL problem mainly depends on forward propagation, which can be given by 
$O\left( KN{{n}_{1}}+K\tau 
{{n}_{L-1}}+\sum\limits_{l=2}^{L-1}{{{n}_{l-1}}{{n}_{l}}} \right)$ and it 
indeed reduces the complexity than NSGA-II.

\section{Numerical Results}
In this section, we use Monte Carlo simulation to verify the closed-form
expression derived. With the numerical result, we analyze the SE of unicast and
multigroup multicast users under different precoding schemes.
\subsection{Simulation Parameters}
 \begin{figure}
	\centering
	 \includegraphics[width=8.5cm]{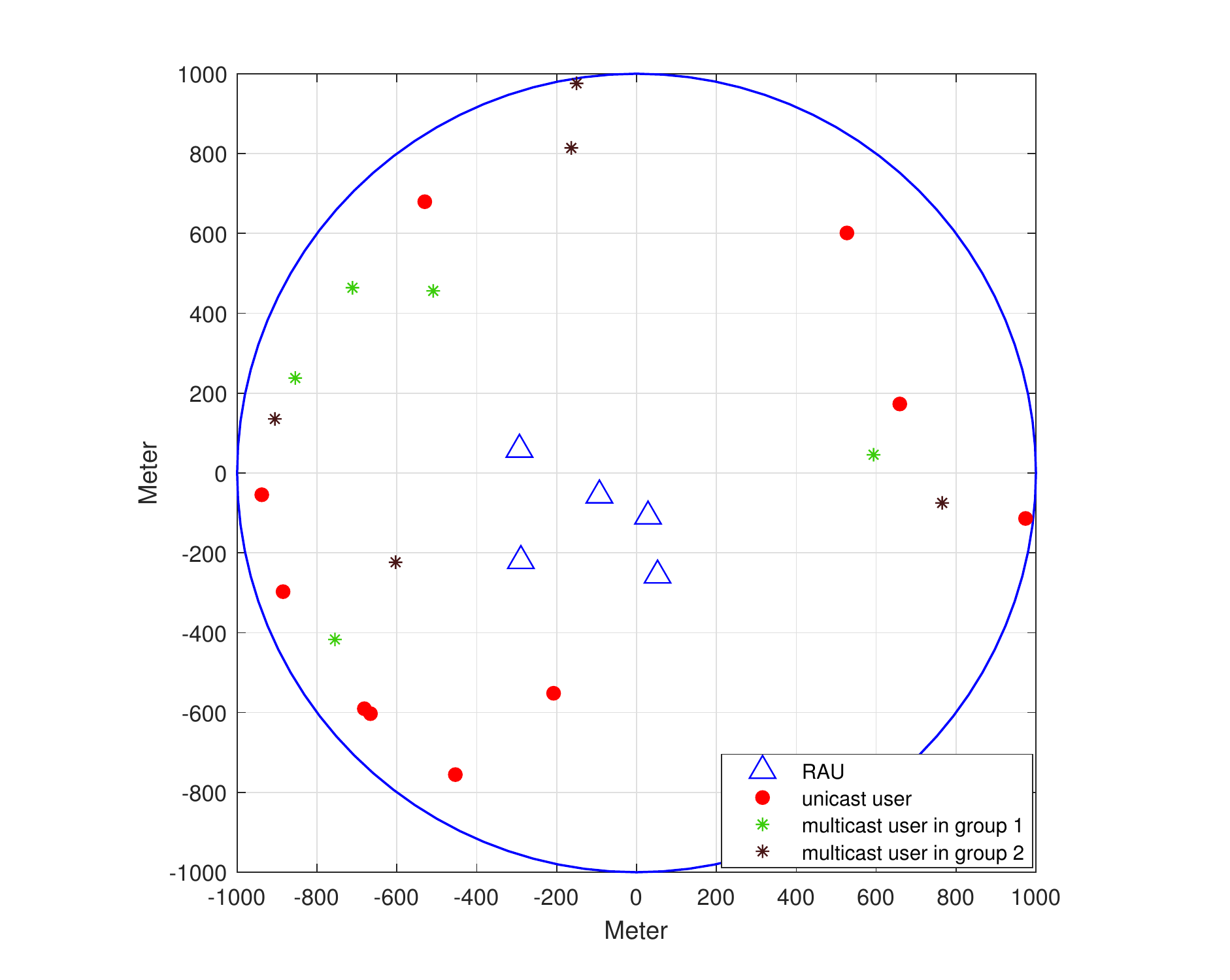}
	\caption{{Simulation system configure.}\label{fig2}}
\end{figure}

\begin{table}
	\small
	\centering
	\caption{Simulation parameter settings}\label{tab1}
	\begin{tabular}{cccccccccccc}
		\hline
		Item& Number
		\\
		\hline
		Number of unicast users $U$    &  10 \\
		\hline
		Number of multicast groups $M$    & 2 \\
		\hline
		Number of multicast users in each group $K_m$     & [5,5] \\
		\hline
		Path loss index $a $            & 3.7 \\
		\hline
		Uplink power  $p_{\mathrm{ul},u}$ and	 ${q}_{\mathrm{ul},m,k}$     &
		[0.5w, 0.5w]\\
		\hline
		Downlink power $p_{\mathrm{dl},u}$ and ${q}_{\mathrm{dl},m}$    &
		[1w, 0.5w] \\
		\hline
		Noise power        & -70 dBm\\
		\hline
		Coherence time T      & 196\\
		\hline
	\end{tabular}
\end{table}
As it is shown in Fig. \ref{fig2}, in simulation, assuming in a curricular area,
there are 5 RAUs and 20 user terminals
randomly distributed. The path loss for unicast and multcast users are defined
as ${{\beta}_{n,u}}=bd_{n,u}^{-a }$ and ${{\eta }_{n,m,k}}=bd_{n,m,k}^{-a }$
respectively, where $d$ is the distance between RAUs and users, $a$ is path
loss index and $b$ is the median of the average path gain at the reference
distance $d=1\text{km}$. The specific parameters are shown in Table \ref{tab1}.
\subsection{Simulation Result Analysis}
\begin{figure}
	\begin{minipage}[t]{0.48\textwidth}
	\centering
	 \includegraphics[width=8.5cm]{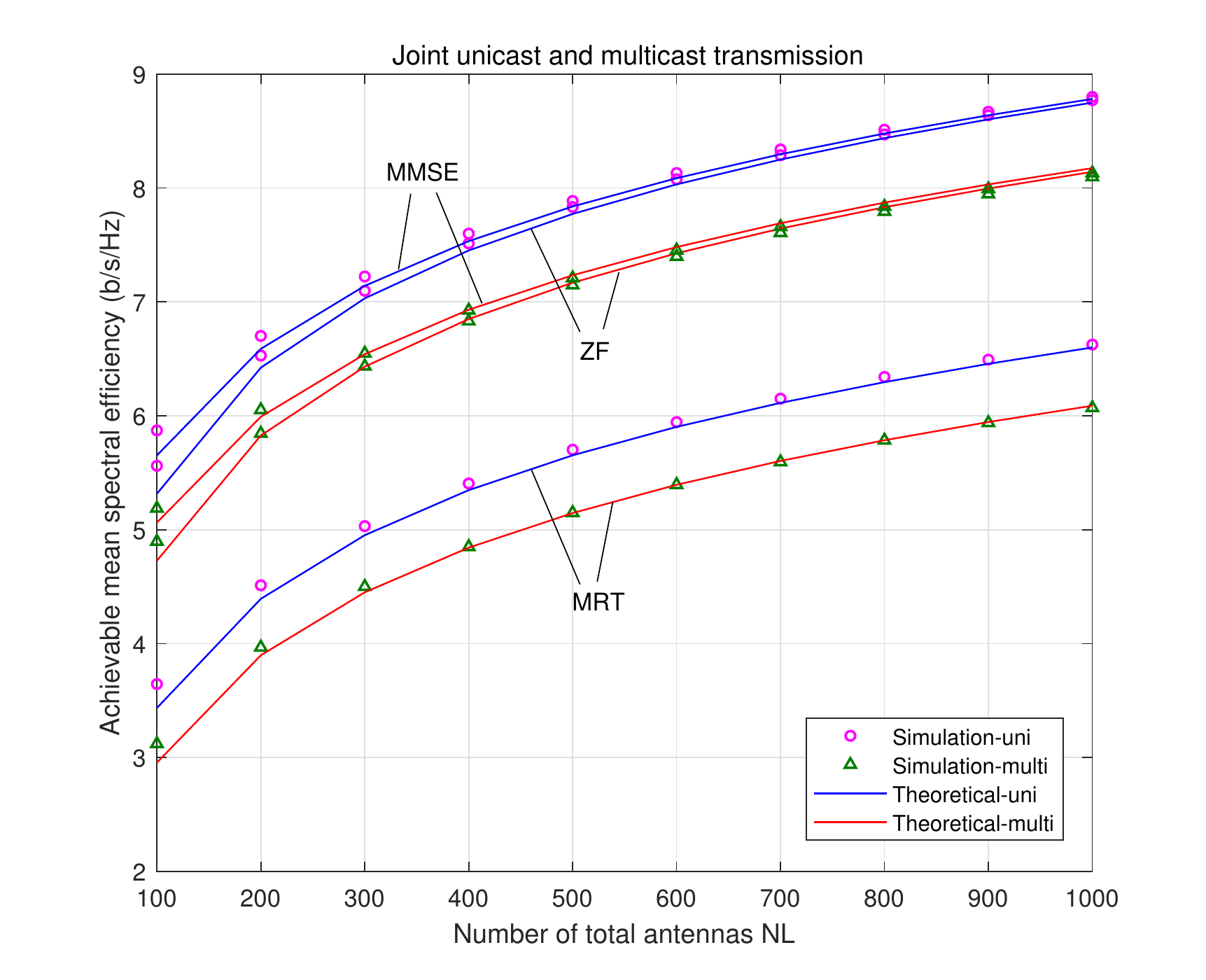}
	\caption{{Spectral efficiency against the total number of
			antennas}\label{fig3}}
	\end{minipage}
    \begin{minipage}[t]{0.48\textwidth}
    		\centering
    	\includegraphics[width=8.5cm]{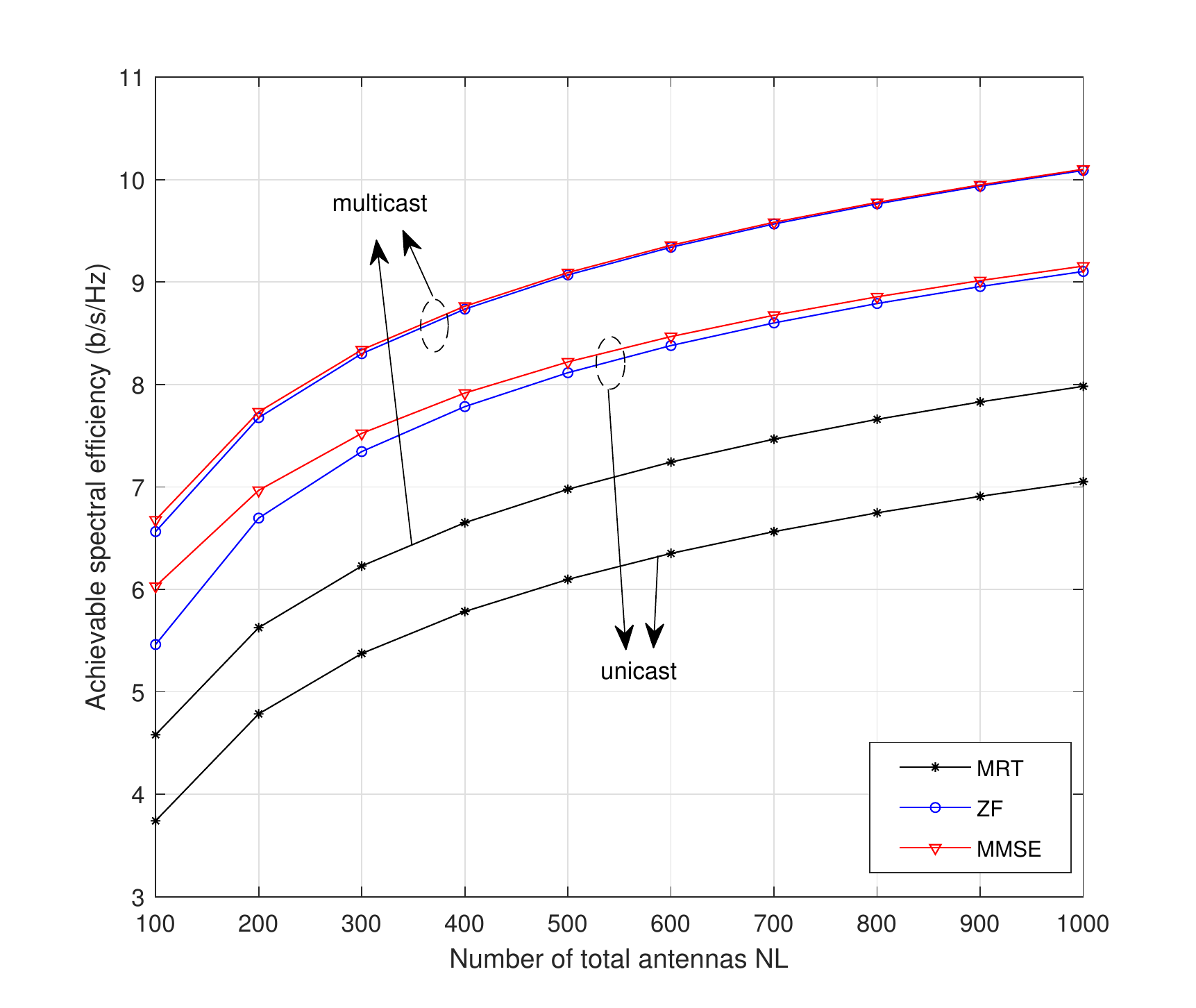}
    	\caption{{Comparison between multicast and unicast}\label{fig4}}
    \end{minipage}
\end{figure}

Fig. \ref{fig3} verifies the theoretical and simulation values of the
achievable SE of unicast users and multigroup multicast users with MRT, ZF and
MMSE precoding against the number of total antennas. It is illustrated that the
simulation results agree well with the theoretical results we deduced, which
verifies the accuracy of \eqref{closeun} and
\eqref{closemu}.

Since both RAUs and multigroup multicast users are dispersedly 
distributed in cell-free distributed massive MIMO scenarios. The distance from 
each multicast group to the RAUs is more even. Thus multigroup multicast users 
are less susceptible than unicast by the location of RAUs.

Fig. \ref{fig4} illustrates the comparison of SE between unicast and multicast
under the same parameters. We have twenty users in each mode and in multicast,
the users are divided into four groups with five users in each group. To
guarantee the fairness of the comparison, the user positions are set the same.
The downlink transmitted power is set to 2 W for each multigroup and unicast
user. However, as shown in Fig. \ref{fig3}, the achievable SE of multicast
users is obviously higher than that of unicast users. It shows that in the same
condition, the power utilization of multicast is higher than unicast and it can
significantly increase the SE when the number of users in a multicast group is
big.

%

\begin{figure}[t]
	\begin{minipage}[t]{0.48\textwidth}
				\centering
				 \includegraphics[width=8.5cm,height=7.2cm]{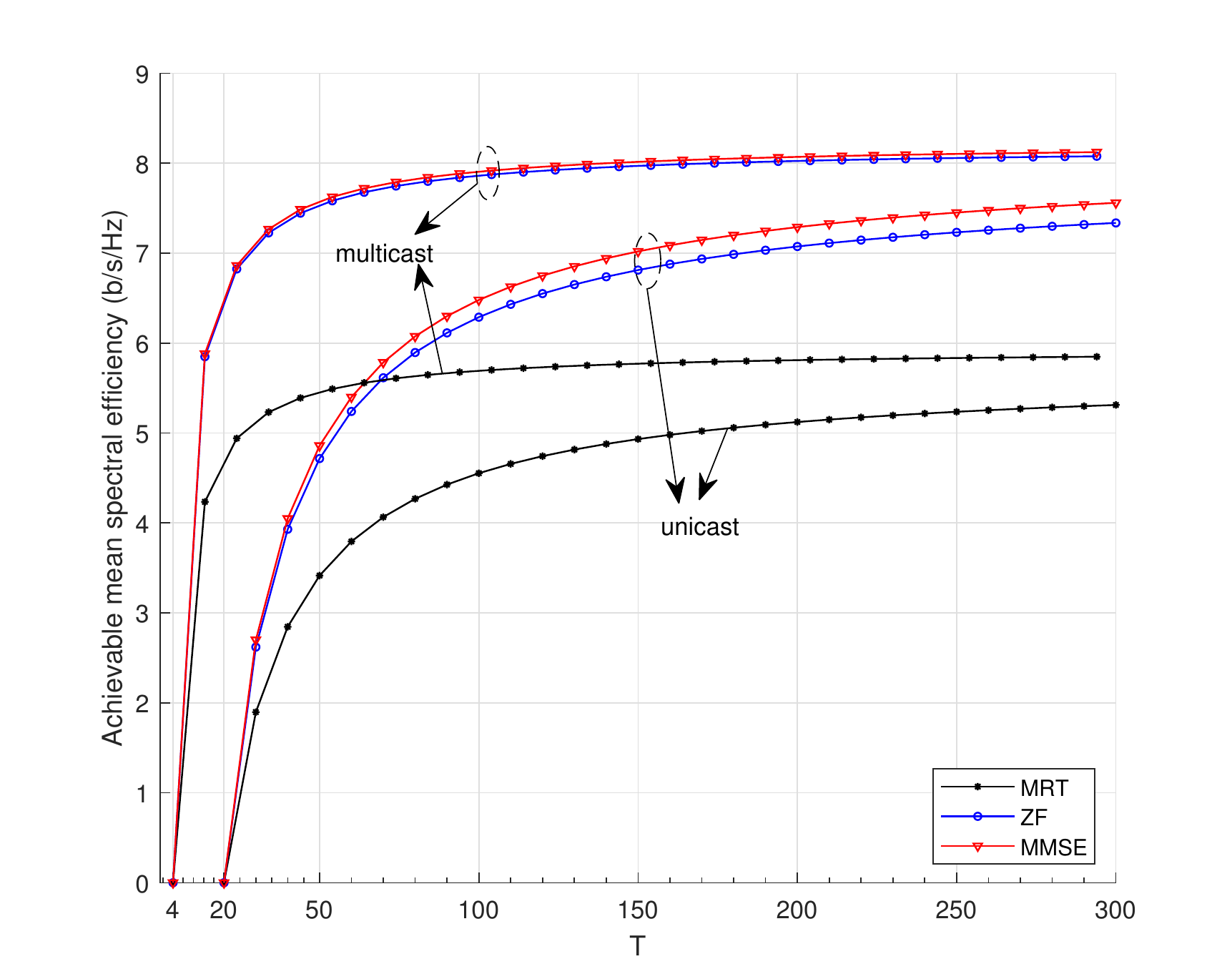}
				\caption{{Spectral efficiency against the coherence time
				T}\label{fig5}}
\end{minipage}
	\begin{minipage}[t]{0.48\textwidth}
	\centering
	 \includegraphics[width=8.5cm,height=7.2cm]{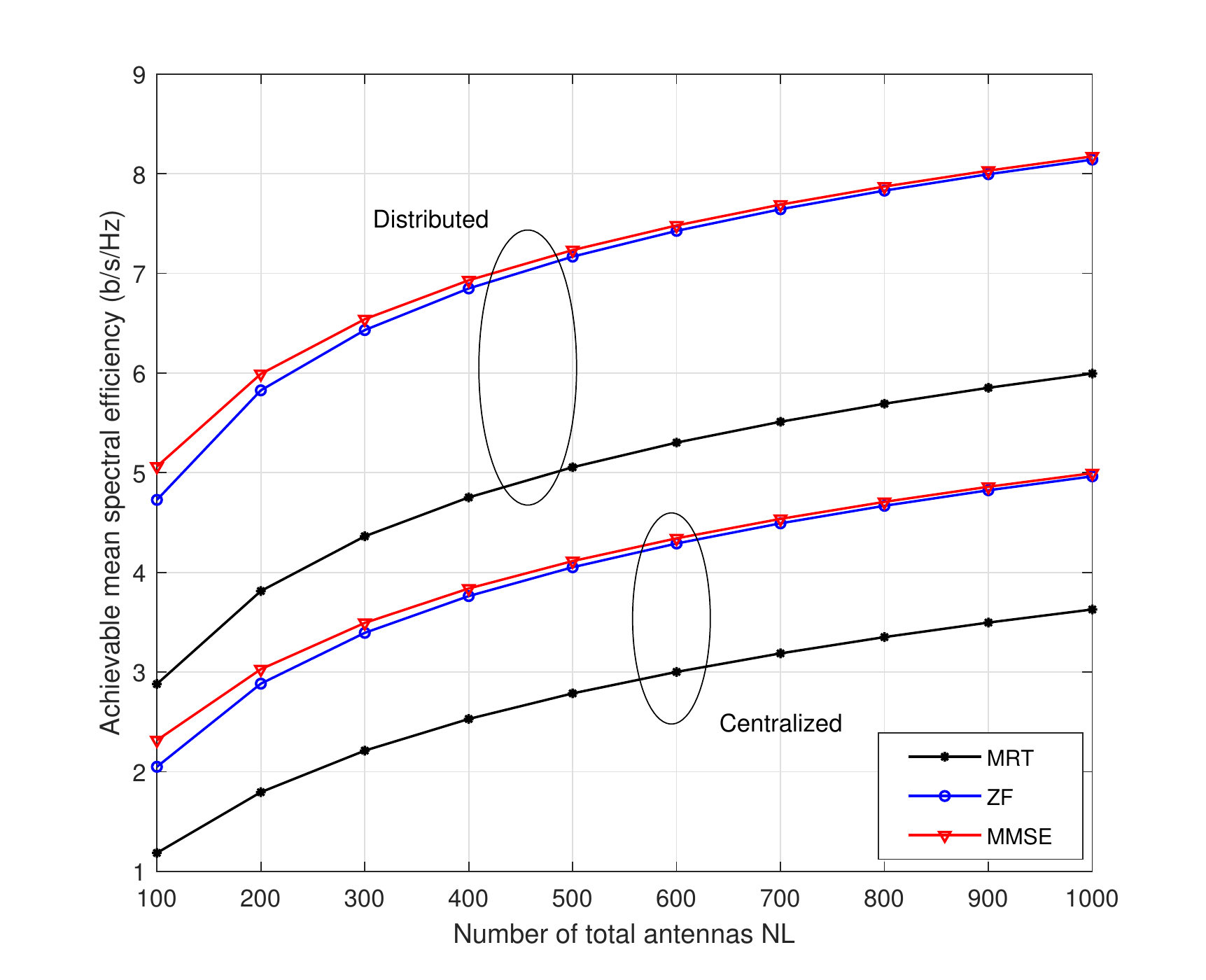}
	\caption{{Comparison of multicast between distributed massive MIMO and
			centralized massive MIMO}\label{fig7}}
	\end{minipage}
\end{figure}

Fig. \ref{fig5} compares the SE of unicast and multicast against the coherence
time $T$. The number of antennas per RAU is set to 50 and the coherence time
$T$ is ranging from the number of pilots $\tau$ to 300. To avoid pilot
contamination, it needs to be bigger than the number of unicast users or
multicast groups. As shown in the figure, when $T$ is small, especially when
$T$ is below 50, the achievable SE of multicast user is extremely higher than
that of unicast user. This is because the demand of pilot cost is much lower in
multicast. As a result, the advantage of multicast in reducing pilot cost can
effectively enhance the SE of users  especially when the coherence time $T$ is
not relatively high.

Fig. \ref{fig7} compares the SE of multicast in cell-free
distributed massive MIMO and traditional centralized massive MIMO systems.
The validity of multicast in centralized massive MIMO has already been
indicated in \cite{sadeghi2018joint}, thus the SE of multicast under the two
scenarios can be compared. The position of users, the number of total
antennas and the other parameters are set exactly the same.
It can be seen from the figure that the SE of multicast users in
cell-free distributed massive MIMO is twice over that in centralized massive
MIMO systems. Actually, it should be indicated that the performance of
multicast in cell-free scenarios to some extent depends on the distribution
of the RAUs and the multicast users.
However, in most times it can achieve better performance due to the
interactions between RAUs and the dispersed distribution of multicast users.
As a result, combining cell-free scenarios with multicast can effectively
improve the spectrum utilization of users.

\begin{figure}[t]
	\begin{minipage}[t]{0.50\textwidth}
	\centering
	\subfigure[noise power -10dbm.]{ \label{fig_5a}
		\includegraphics[width=9cm]{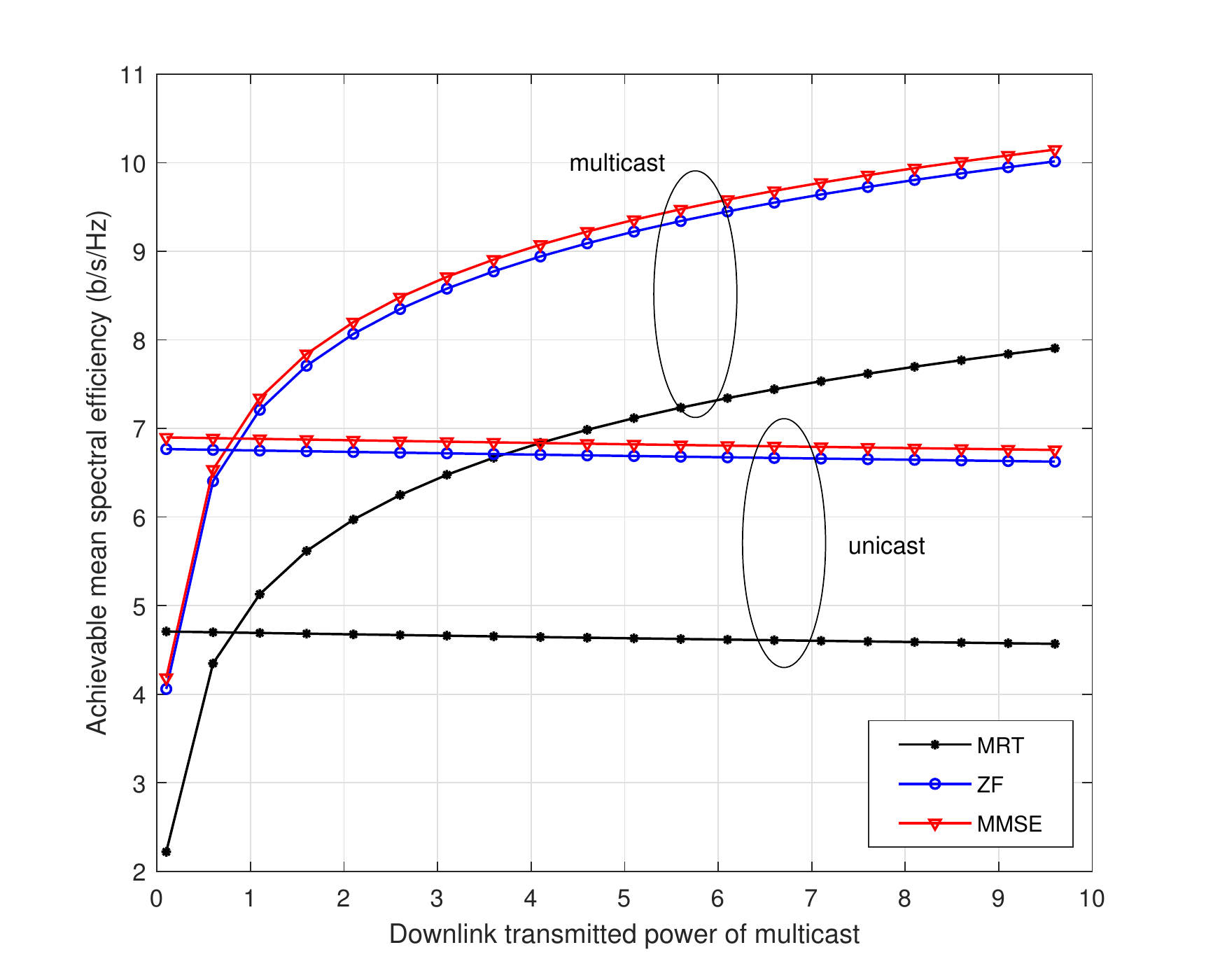} }
		\end{minipage}
		\begin{minipage}[t]{0.50\textwidth}
	\subfigure[{noise power -15dbm.}] {
		\label{fig_5b}
		\includegraphics[width=9cm]{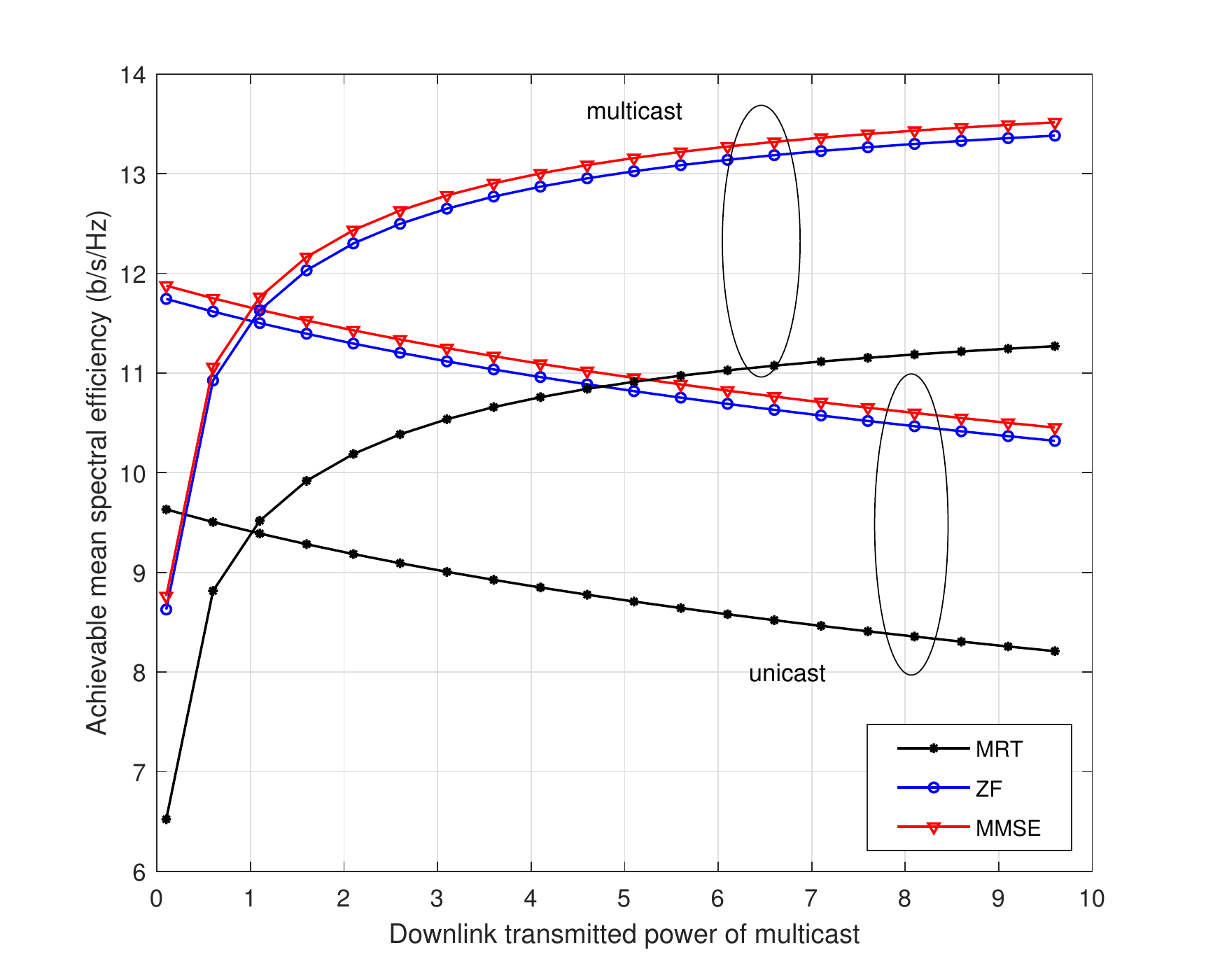}}
		\end{minipage}
	\caption{Spectral efficiency against downlink transmitted power of
		multicast.}
	\label{fig6}
\end{figure}

Fig. \ref{fig6} illustrates the SE of unicast and multicast users against
the downlink transmitted power of multicast. {The number of 
antennas per RAU is set to 50 and} it is assumed that the downlink
transmitted power of multicast ranging from 0.1 W to 9.6 W and the power of
unicast remains 1 W.
Fig. \ref{fig_5a} and Fig. \ref{fig_5b} are the results under the noise power
of -70 dBm and -120dbm respectively. The rest simulation parameters are set in
accordance with Table \ref{tab1}.
It can be seen from the two figures that the SE of multicast users
increases with the increasing downlink transmitted power of multicast groups,
while the SE of unicast users are affected mildly and dropped slightly
especially in  Fig. \ref{fig_5a}.
As a result, in joint unicast and multigroup multicast transmission, increasing
the transmitted power of multicast can promote the SE of multicast and not
cause a big trouble to unicast. This supports the feasibility of the joint
unicast and multicast system proposed in this paper. Besides, under the
condition of lower noise power, which
means high signal-to-noise ratio (SNR), the downlink 
transmitted power of
multicast has a greater impact on unicast. This is because both unicast and
multicast are more sensitive to the interferences between them in high SNR
cases. In these cases, the confliction between maximizing the SE of unicast and
multicast are more obvious, which leads to the MOOP proposed in \eqref{moop}.

\begin{figure}
		\begin{minipage}[t]{0.48\textwidth}
	\centering
	 \includegraphics[width=8.5cm]{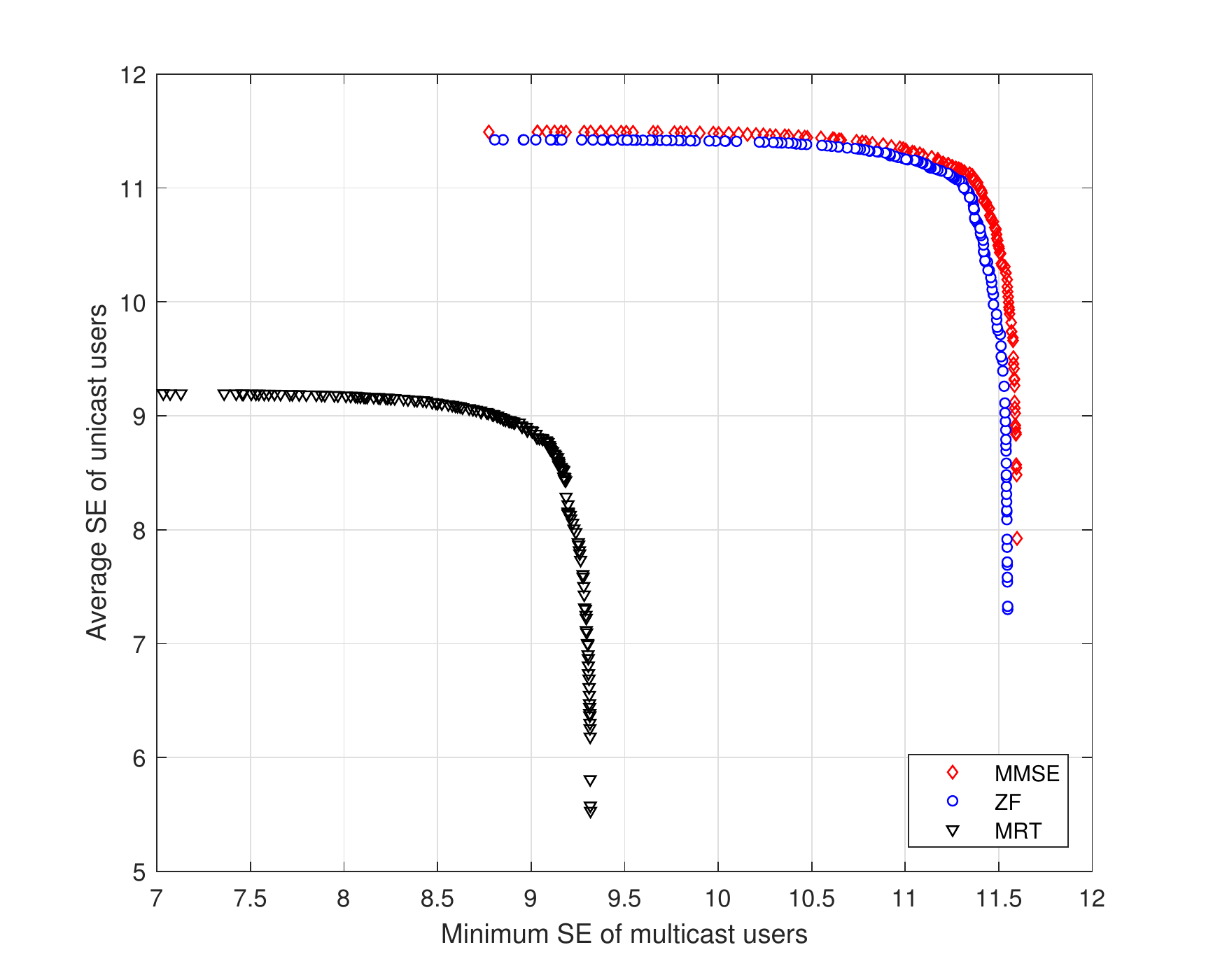}
	\caption{{Trade-off between the minimum SE of multicast users and average
	SE of unicast users }\label{fig8}}
\end{minipage}
	\begin{minipage}[t]{0.48\textwidth}
			\centering
		\includegraphics[width=8.5cm]{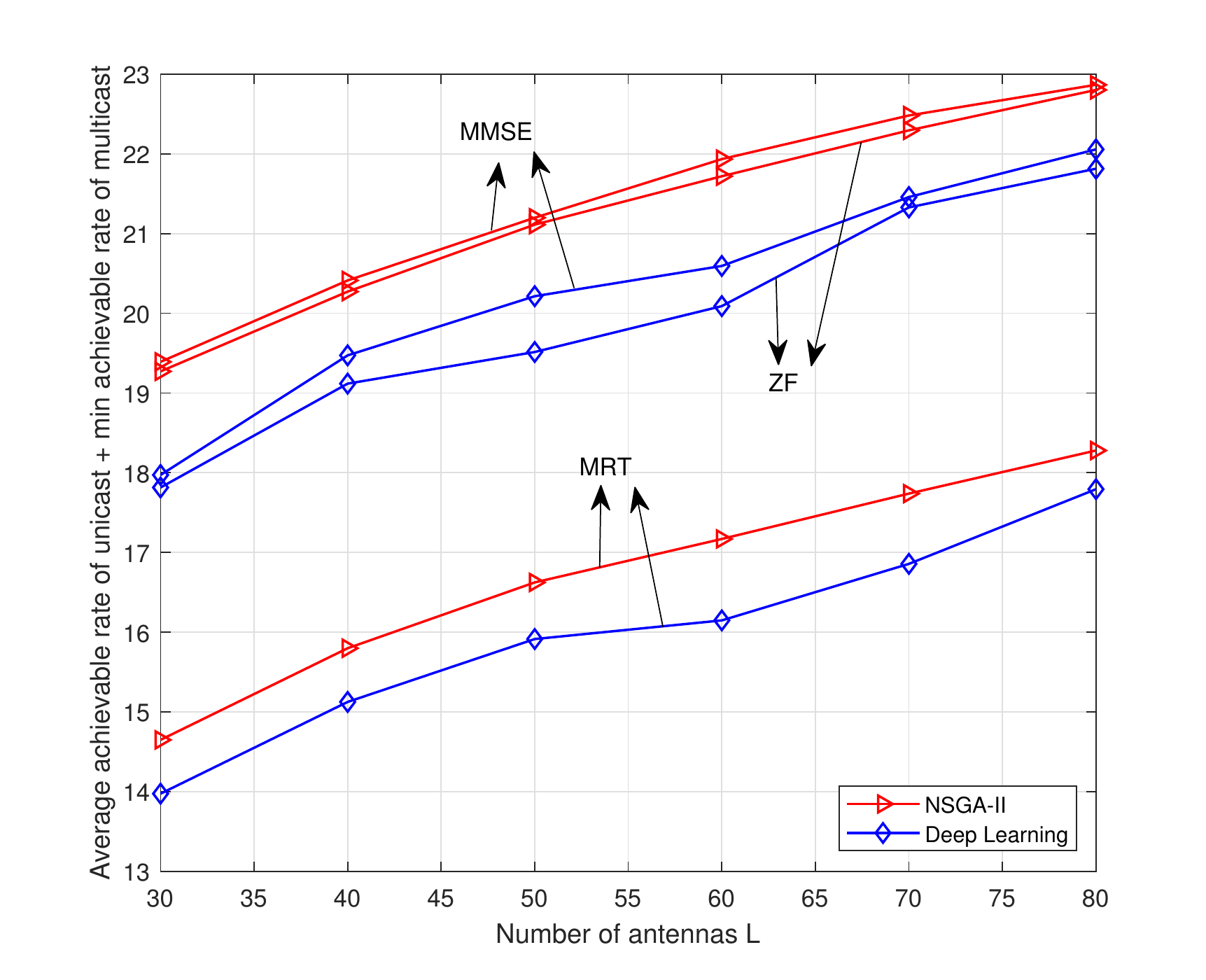}
		\caption{{Comparison of the sum achieveable rate between NSGA-II and
		Deep
				Learning
			}\label{fig9}}
	\end{minipage}
\end{figure}
\begin{figure}
		\begin{minipage}[t]{0.48\textwidth}
		\centering
		\includegraphics[width=8.5cm]{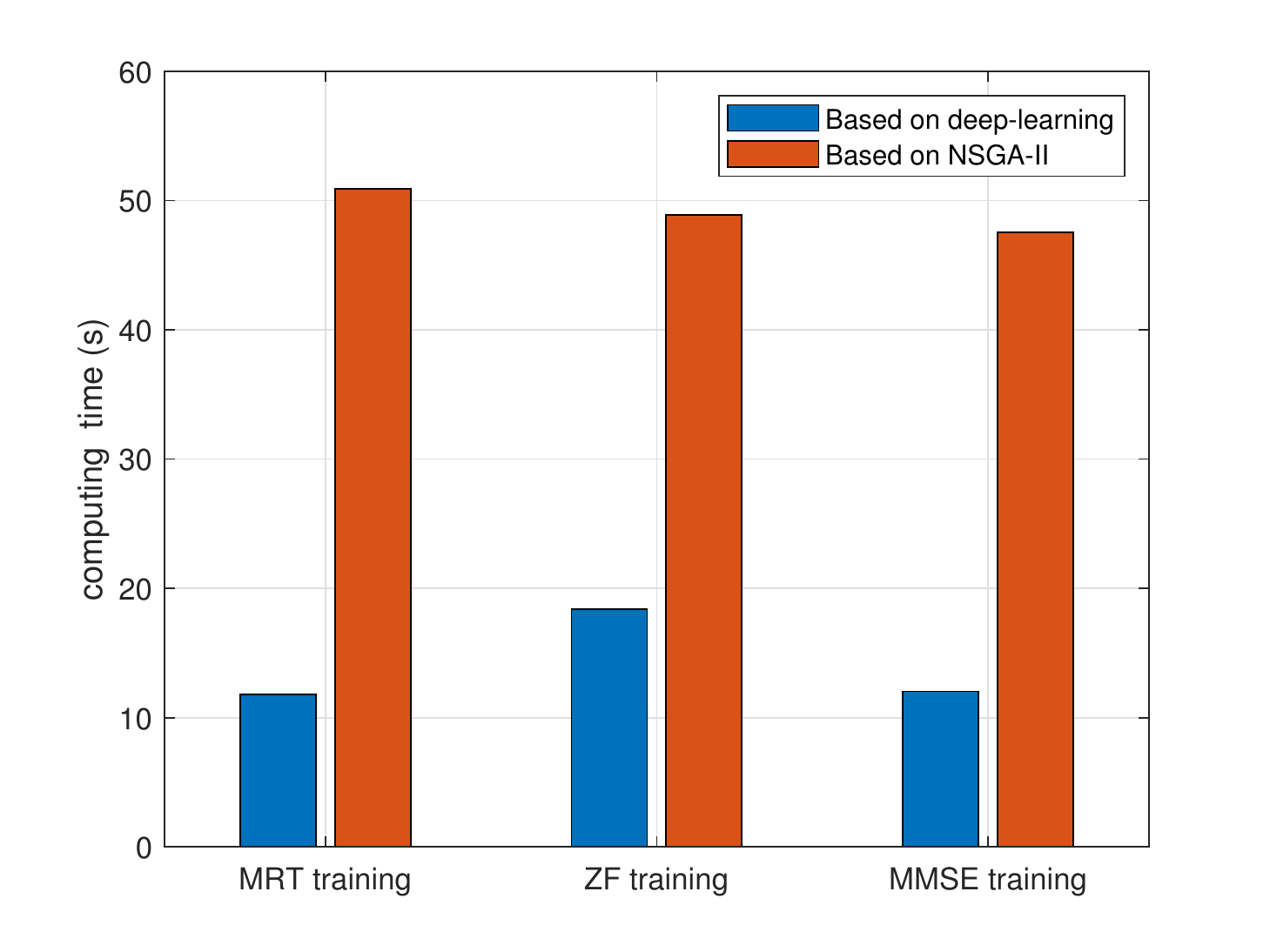}
		\caption{{Comparison of elapsed time between NSGA-II and
				Deep
				Learning
			}\label{fig10}}
	\end{minipage}
\end{figure}

Fig. \ref{fig8} illustrates the trade-off between the minimum SE of multicast
users and average SE of unicast users under different precoding schemes. The
picture shows the Pareto boundary obtained by solving the MOOP \eqref{moop}
with NSGA-II. The number of antennas per RAU is set to 50. {The 
upper limit of downlink transmitted power per RAU is given by 10W.} 
The required minimum SE of both unicast and multicast users is set to 3 b/s/Hz.
It can be seen that the minimum SE of multicast users increases with the
decrease of the average SE of unicast users, which means that they can not
achieve the highest value simultaneously and verifys the confliction between
these two objectives in joint unicast and multigroup multicast transmission
under cell-free distributed massive MIMO scenarios. As is marked in the
picture, the approximately optimal value of the trade-off between multicast and
unicast can be obtained by NSGA-II, which is also the optimal solution of the
MOOP \eqref{moop}.

Fig. \ref{fig9} illustrates the comparison of the maximum
sum of average achievable rate of unicast users and minimum achievable rate of
multicast users between	NSGA-II and DL. We input the channel large-scale
fading between 5 RAUs and 4 unicast users and 4 multicast users evenly divided
into 2 multicast group to the DNN and the corresponding parameters in NSGA-II
is set the same. To adapt the input environment of DL, the 
$2\times 2$ two-dimensional
matrix and $5\times 2\times 2$ three-dimensional matrix need to be merged into 
one matrix and
reshaped into a $1\times 40$ column vector. The total power for downlink 
transmission
are 47 W. To set a higher standard, we select the best point of the sum
rate in	NSGA-II for comparison. It can be seen from the figure that the sum
rate obtained by DL can achieve above 92$\%$ of the highest rate obtained by
NSGA-II. {Fig. \ref{fig10} shows the elapsed time of DL and 
NSGA-II with $L=60$.}  {In DL, it only takes about 14 seconds 
to train, 
however, with NSGA-II, it takes about 50 seconds, which is above 3.5 times of 
the elapsed time with DL.} The effect of NSGA-II algorithm optimization is 
quite good, but the 	
calculation time is long and the complexity is high. In DL, the optimization
method is to train the loss function to the lowest, so the result will be
affected by some neural network parameters and the performance will not achieve
the best, but the optimization time of DL is quite short and it is considerable 
for DL to be
close to NSGA-II.

\section{Conclusion}
In this paper, we studied the joint unicast and multigroup multicast
transmission in cell-free distributed massive MIMO systems. With the estimated
CSI obtained by co-pilot assignment strategy,
we deduced the closed-form expression of downlink SE of unicast users and
multicast users with MRT, ZF and MMSE precoding schemes.
With these expressions, we proposed a MOOP between the MMF of
multicast and the maximum of the average SE of unicast under several
constraints. Based on the MOOP, we further converted it into a DL
problem to reduce computational complexity.
We verified the derived closed-form expressions by Monte Carlo simulation and
compared the SE of unicast and multicast under the same parameters to show that
multicast can achieve higher SE and occupy less coherence time slots.
Besides, it was verified that the combination of distributed massive MIMO
systems and multicast transmission can effectively achieve higher SE. The
effects of downlink transmitted power and noise power on SE were analyzed and
the two objectives in the proposed MOOP was proved conflicting.
The trade-off region between these two conflicting problems obtained with
NSGA-II under various constraints offered numerous flexibilities for system
optimization. Furthermore, the comparison between NSGA-II and DL showed that
solving the MOOP by DL can also achieve relatively good results with quite
short elapsed time and low complexity.

\appendices
\section{Proof of Theorem 1}\label{ap}
\setcounter{equation}{0}
\renewcommand\theequation{A\arabic{equation}}
We first calculate the needed signal for unicast users (MMSE as an example).
By substituting the MMSE precoding vector into the item
$\mathbb{E}\left[\mathbf{c}_{k}^{\mathrm{H}}\mathbf{v}_{{k}}^{\mathrm{pre}}
\right]$ , we have
\begin{align}
\notag	\mathbb{E}\left[
	\mathbf{c}_{k}^{\mathrm{H}}\mathbf{v}_{{k}}^{\mathrm{MMSE}}\right]
&=\mathbb{E}\left[
	\mathbf{c}_{k}^{\mathrm{H}}\sqrt{\frac{p_{\text{dl},k}
			{{(A{{\theta}_{k}}+\sigma _{\mathrm{dl}}^{2})}^{2}}}
		{A{{\theta}_{k}}}}{{\chi}_{k}} \right] \\
	& =\sqrt{\frac{p_{\mathrm{dl},k}{{(A{{\theta}_{k}}
				+\sigma_{\mathrm{dl}}^{2})}^{2}}}{A{{\theta }_{k}}}}
	{{(A{{\theta}_{k}}+\sigma _{\mathrm{dl}}^{2})}^{-1}}NL{{\theta
	}_{k}}\nonumber\\
 &=\sqrt{\frac{{{\left( NL \right)}^{2}}p_{\mathrm{dl},k}{{\theta}_{k}}}{A}}  ,
\end{align}
where ${{\chi }_{k}}=\left[\mathbf{\hat{Q}}{{(\overset{{}}
		{\mathop{{{{\mathbf{\hat{Q}}}}^{\mathrm{H}}}\mathbf{\hat{Q}+}
				\sigma	_{\mathrm{dl}}^{2}}}\,\otimes
		{\mathbf{I}_{NL}})}^{-1}}\right]_{k}$.
	
Then, for the denominator part $\mathbb{E}\left[ {{\left|
		\mathbf{c}_{k}^{\mathrm{H}}\mathbf{v}_{u}^{\mathrm{MMSE}} \right|}^{2}}
\right]$,
when $u=k$, due to the independence of $\mathbf{c}_{k}$ and
	${\mathbf{n}}_{k}$, we have $\mathbb{E}\left[
	\mathbf{c}_{k}{{\mathbf{n}}_{k}} \right]=0 $, and we can obtain
\begin{align}
\notag& \mathbb{E}\left[ {{\left|
			\mathbf{c}_{k}^{\mathrm{H}}\mathbf{v}_{k}^{\mathrm{MMSE}}
			\right|}^{2}}
	\right] \nonumber\\
=&\,\frac{p_{\mathrm{dl},k}{{(A{{\theta
					}_{k}}+\sigma _{\mathrm{dl}}^{2})}^{2}}}{A{{\theta
			}_{k}}}\mathbb{E}\left[ \left(
			\mathbf{c}_{k}^{\mathrm{H}}{{\chi }_{k}}
	\right){{\left( {{\chi }_{k}^{\mathrm{H}}}\mathbf{c}_{k}
			\right)}} \right] \nonumber\\
\notag	=&\,p_{\mathrm{dl},k}{{\left( A{{\theta }_{k}}
			\right)}^{-1}}\mathbb{E}\left[
	\mathbf{c}_{k}^{\mathrm{H}}\mathbf{\hat{c}}_{k}^{{}}
	\mathbf{\hat{c}}_{k}^{\mathrm{H}}\mathbf{c}_{k}^{{}}
	\right] \\
\notag	 \overset{(a)}{=}&\,\frac{p_{\mathrm{dl},k}\mathbb{E}\left[
		\mathbf{c}_{k}^{\mathrm{H}}(\sqrt{\tau
			p_{\mathrm{ul},k}}{{\mathbf{c}}_{k}}+{{\mathbf{n}}_{k}})
		{{(\sqrt{\tau p_{\mathrm{ul},k}}{{\mathbf{c}}_{k}}+{{\mathbf{n}}_{k}})
			}^{\mathrm{H}}}\mathbf{c}_{k}^{{}} \right]}{A\left( \tau
		p_{\mathrm{ul},k}\sum\limits_{n=1}^{N}{{{\beta }_{n,k}}}+\sigma
		_{\mathrm{ul}}^{2} \right)} \\
\notag	\overset{(b)}{=}&\,\frac{p_{\mathrm{dl},k}\left( \tau
	p_{\mathrm{ul},k}{{\left(
				NL\sum\limits_{n=1}^{N}{{{\beta }_{n,k}}} \right)}^{2}}
		\right)}{A\left( \tau
		p_{\mathrm{ul},k}\sum\limits_{n=1}^{N}{{{\beta }_{n,k}}}+\sigma
		_{\mathrm{ul}}^{2}
		\right)}+\frac{1}{NL}p_{\mathrm{dl},k}\sum\limits_{n=1}^{N}{{{\beta
		}_{n,k}}} \\
=&\,\frac{{{\left( NL \right)}^{2}}p_{\mathrm{dl},k}{{\theta
			 }_{k}}}{A}+\frac{1}{NL}p_{\mathrm{dl},k}\sum\limits_{n=1}^{N}{{{\beta
			}_{n,k}}}  ,
\end{align}
where (a) is obtained by plugging $\theta_{k}$ into \eqref{fkjc} and (b)
is got by means of uncorrelated vectors and the large-scale
random matrix conclusions\cite{wang2019performance}.

{\emph{Lemma (\cite{wang2019performance})}: If $\mathbf{A}\in {{C}^{M\times 
			M}}$ 
	has 
	the uniformly bounded spectral norm (relative to $M$), vector x and y 
	follows 
	$\mathcal{C}\mathcal{N}\left( 0,\frac{1}{M}{{\mathbf{I}}_{M}} \right)$, and 
	x, 
	y is independent of each other and matrix A, it has
	\begin{align}
		\notag		& {{\mathbf{x}}^{\mathrm{H}}}\mathbf{Ay}\xrightarrow[M\to 
		\infty 
		]{a.s.}0 
		\\ 
		\notag		& 
		{{\mathbf{x}}^{\mathrm{H}}}\mathbf{Ax}-\frac{1}{M}\mathrm{Tr}\left( 
		\mathbf{A} 
		\right)\xrightarrow[M\to \infty ]{a.s.}0 
	\end{align}
}

When $u\ne k$,
\begin{align}
 \mathbb{E}\left[ {{\left|
	\mathbf{c}_{k}^{\mathrm{H}}\mathbf{v}_{u}^{\mathrm{MMSE}}
	\right|}^{2}} \right]
 &=\frac{1}{NL}p_{\mathrm{dl},u}\sum\limits_{n=1}^{N}{{{\beta }_{n,k}}}.
\end{align}

For the denominator part $\mathbb{E}\left[ {{\left|
		\mathbf{c}_{k}^{\mathrm{H}}\mathbf{w}_{r}^{\mathrm{MMSE}} \right|}^{2}}
\right]$, based on large-scale
random matrix theory, we have
\begin{align}
\mathbb{E}\left[ {{\left|
	\mathbf{c}_{k}^{\mathrm{H}}\mathbf{w}_{r}^{\mathrm{MMSE}}
	\right|}^{2}}\right]
&=\frac{1}{NL}q_{\mathrm{dl},r}\sum\limits_{n=1}^{N}{{{\beta }_{n,k}}}.
\end{align}
The conclusion can be proved by substituting the values of the above fractions
into \eqref{sinrun}.

For multicast, we can calculate the needed signal for user
(Molecularpart) similarly, by substituting the preocoding vector, we have
\begin{align}
\notag
&\mathbb{E}\left[\mathbf{t}_{m,k}^{\mathrm{H}}\mathbf{w}_{m}^{\mathrm{MMSE}}
\right] \\ \nonumber
=&\,\sqrt{q_{\mathrm{dl},m}\frac{{{(A{{\upsilon}_{m}}
		+\sigma_{\mathrm{dl}}^{2})}^{2}}}{A{{\upsilon}_{m}}}}
	\mathbb{E}\left[\mathbf{t}_{m,k}^{\mathrm{H}}{{\chi}_{m}}\right] \nonumber\\
\notag	\overset{\text{(a)}}=&\,
\sqrt{q_{\mathrm{dl},m}\frac{{{(A{{\upsilon}_{m}}
		+\sigma_{\mathrm{dl}}^{2})}^{2}}}{A{{\upsilon}_{m}}}}
		\frac{NL\sum\limits_{n=1}^{N}{{\kappa }_{n,m}}
			\sqrt{\tau q_{\mathrm{ul},m,k}}{\eta_{n,m,k}}}
		{{{(A{{\upsilon}_{m}}+\sigma_{\mathrm{dl}}^{2})}}} \nonumber \\
{=}&\,\sqrt{\frac{{{\left( NL \right)}^{2}}q_{\mathrm{dl},m}{{\zeta}_{m,k}}}A},
\end{align}
where 	${{\chi}_{m}}=
\left[\mathbf{\hat{Q}}{{\left({\mathop{{{{\mathbf{\hat{Q}}}}^{\mathrm{H}}}
	\mathbf{\hat{Q}+}\sigma_{\mathrm{dl}}^{2}}}\,\otimes
	{\mathbf{I}_{NL}}\right)}^{-1}}\right]_{U+m}, {{\kappa}_{n,m}}=$ 
	${\sum\nolimits_{j=1}^{{{K}_{m}}}{\tau
		q_{\mathrm{ul},m,j}\eta_{n,m,j}}}/\left({\sum\nolimits_{j=1}^{{{K}_{m}}}
	{\tau q_{\mathrm{ul},m,j}\eta _{n,m,j}}+\sigma_{\mathrm{ul}}^{2}}\right), 
	{{\zeta 
	}_{m,k}} =\sum_{n=1}^{N}{\frac{\tau
		q_{\mathrm{ul},m,k}\eta_{n,m,k}^{2}}
		{\sum\nolimits_{j=1}^{{{K}_{m}}}{\tau
		q_{\mathrm{ul},m,j}{{\eta}_{n,m,j}}}
		+\sigma_{\mathrm{ul}}^{2}}}$,
(a) is obtained by large-scale random matrix conclusions
\cite{wang2019performance}.

For the denominator part $\mathbb{E}\left[ {{\left|
		\mathbf{t}_{m,k}^{\mathrm{H}}{{\mathbf{w}}_{r}^{\mathrm{MMSE}} }
		\right|}^{2}}
		\right]$, when $r=m$, we have
\begin{align}\label{a2}
\nonumber		& \mathbb{E}\left[ {{\left|	
	\mathbf{t}_{m,k}^{\mathrm{H}}\mathbf{w}_{m}^{\mathrm{MMSE}}\right|}^{2}}
		\right]	\\
	= &q_{\mathrm{dl},m}	
	\frac{{{(A{{\upsilon}_{m}}+\sigma_{\mathrm{dl}}^{2})}^{2}}}
	{A{{\upsilon}_{m}}}\mathbb{E}\left[ \left(
	\mathbf{t}_{m,k}^{\mathrm{H}}{{\chi }_{m}}\right)
	{{\chi}_{m}}^{\mathrm{H}}\mathbf{t}_{m,k}^{{}} \right].
\end{align}
				
Then, for the term $\mathbb{E}\left[
\left(\mathbf{t}_{m,k}^{\mathrm{H}}{{\chi }_{m}}
\right){{\chi}_{m}}^{\mathrm{H}}\mathbf{t}_{m,k}^{{}} \right]$,
due to the large-scale random matrix theory	\cite{wang2019performance},
we have
\begin{align}
\notag&	\mathbb{E}\left[ \left(
		\mathbf{t}_{m,k}^{\mathrm{H}}{{\chi }_{m}}\right)
		{{\chi}_{m}}^{\mathrm{H}}\mathbf{t}_{m,k}^{{}} \right]\\
\notag	=&\, \frac{\mathbb{E}\Big[  \mathbf{t}_{m,k}^{\mathrm{H}}
		\big[ \mathbf{\hat{Q}}\big]_{U+m}
		\Big(\big[ \mathbf{\hat{Q}}\big]_{U+m}	
		\Big)^{\mathrm{H}}\mathbf{t}_{m,k}
		\Big]}{(A{{\upsilon}_{m}}+\sigma_{\mathrm{dl}}^{2})^2}\\
	 =&\,\frac{\sum\limits_{n=1}^{N}{{\kappa}_{n,m}}tr\left(
		 \tau q_{\mathrm{ul},m,k}\sum\nolimits_{j=1}^{{{K}_{m}}}
		 \mathcal{I}_{gj,m,k}
	+\mathcal{I}_{gn,m,k}\right)}{(A{{\upsilon}_{m}}+\sigma_{\mathrm{dl}}^{2})^2}.
	\end{align}

Based on the above analysis and combining the result into
(\ref{a2}), we obtain
\begin{align}
\notag	&	\mathbb{E}\left[ {{\left|
\mathbf{t}_{m,k}^{\mathrm{H}}\mathbf{w}_{m}^{\mathrm{MMSE}}\right|}^{2}}
\right]\\
\notag=&\,\frac{q_{\mathrm{dl},m}\Big( \tau q_{\mathrm{ul},m,k}
 {{\big(NL\sum_{n=1}^{N}{\eta_{n,m,k}} \big)}^{2}}
		\Big)}{A\left( \sum_{n=1}^{N}{\left(
			\sum_{j=1}^{{{K}_{m}}}{\tau q_{\mathrm{ul},m,j}
				\eta_{n,m,j}}+\sigma _{\mathrm{ul}}^{2} \right)}
		\right)}\\
\notag &+\frac{1}{NL}q_{\mathrm{dl},m}\sum_{n=1}^{N}{\eta
		_{n,m,k}} \\
=&\, \frac{{{\left( NL \right)}^{2}}q_{\mathrm{dl},m}{\zeta_{m,k}}}
	{A}+\frac{1}{NL}q_{\mathrm{dl},m}\sum\limits_{n=1}^{N}{\eta_{n,m,k}}  .
\end{align}

When $r\ne m$,
\begin{align}
	\mathbb{E}\left[ {{\left|
	\mathbf{t}_{m,k}^{\mathrm{H}}\mathbf{w}_{r}^{\mathrm{MMSE}}
	\right|}^{2}} \right]
&=\frac{1}{NL}q_{\mathrm{dl}}\sum\limits_{n=1}^{N}{\eta _{n,m,k}}.
\end{align}

Then, to the other denominator part
\begin{align}
	\mathbb{E}\left[ {{\left|
	\mathbf{t}_{m,k}^{\mathrm{H}}\mathbf{v}_{u}^{\mathrm{MMSE}}
	\right|}^{2}}\right]
& =\frac{1}{NL}p_{\mathrm{dl},u}\sum\limits_{n=1}^{N}{{{\eta }_{n,m,k}}}.
\end{align}

The conclusion can be proved by combining the results of the above fractions
into \eqref{sinrmu}. The closed-form expression of ZF can be derived similarly
and the derivation of the needed signals with MRT have some difference and is
given below
\begin{align}
\notag&	
\mathbb{E}\left[\mathbf{c}_{k}^{\mathrm{H}}\mathbf{v}_{k}^{\mathrm{MRT}}\right]\\
	\notag	=&\sqrt{\frac{p_{\mathrm{dl},k}}{L{\theta}_{k}}}
	\mathbb{E}\left[\mathbf{c}_{k}^{\mathrm{H}}{{{\mathbf{\hat{c}}}}_{k}}
	\right] \\
	\notag
	=&\,\sqrt{\frac{p_{\mathrm{dl},k}}{L{{\theta}_{k}}}}
	\sum\limits_{n=1}^{N}
	{\frac{\sqrt{\tau p_{\mathrm{ul},k}}{{\beta}_{n,k}}}
		{\sigma_{\mathrm{ul}}^{2}{{\mathbf{I}}_{NL}}\text{+}\tau
			p_{\mathrm{ul},k}
			{{\beta }_{n,k}}}}\left(\begin{aligned}
		&\sqrt{\tau p_{\mathrm{ul},k}}\mathbb{E}\left[\mathbf{c}_{k}
		^{\mathrm{H}}{{\mathbf{c}}_{k}}\right]\\& +\mathbb{E}\left[
		\mathbf{c}_{k}^{\mathrm{H}}{{\mathbf{n}}_{k}}
		\right]\end{aligned}\right) \\
	\notag{\overset{\text{(a)}}=}&\,
	\sqrt{\frac{p_{\mathrm{dl},k}}{L{{\theta
				}_{k}}}}\sum\limits_{n=1}^{N}{\frac{L\tau p_{\mathrm{ul},k}
			{{\beta}_{n,k}}^{2}}
		{\sigma _{\mathrm{ul}}^{2}{{\mathbf{I}}_{NL}}\text{+}\tau
			p_{\mathrm{ul},k}{{\beta }_{n,k}}}} \\
	=&\,\sqrt{Lp_{\mathrm{dl},k}{{\theta }_{k}}},
\end{align}
where (a) is got by the uncorrelated vectors between channels and noises and the
large-scale random matrix conclusions.
\begin{align}
\notag	& \mathbb{E}\left[
	\mathbf{t}_{m,k}^{\mathrm{H}}\mathbf{w}_{m}^{\mathrm{MRT}}
	\right]\\\notag=&\,\sqrt{\frac{q_{\text{dl},m}}{L{{\upsilon  }_{m}}}}
	\mathbb{E}\left[
	\mathbf{t}_{m,k}^{\mathrm{H}}{{{\mathbf{\hat{t}}}}_{m}} \right] \\
\notag	=&\,\sqrt{\frac{q_{\mathrm{dl},m}}{L{{\upsilon}_{m}}}}
		\sum\limits_{n=1}^{N}{{\kappa}_{n,m}}\sqrt{\tau
		q_{\mathrm{ul},m,k}}\mathbb{E}\left[
	\mathbf{t}_{m,k}^{\mathrm{H}}\mathbf{t}_{m,k} \right] \\
\notag	{\overset{\text{(b)}}=}&\,
\sqrt{\frac{q_{\mathrm{dl},m}\left(L\sum_{n=1}^{N}
			{{\kappa}_{n,m}}\sqrt{\tau
				q_{\mathrm{ul},m,k}}L\sum_{n=1}^{N}
			{\eta_{n,m,k}^{{}}}\right)^2}{\sum_{n=1}^{N}
			{\frac{{{(\sum_{j=1}^{{{K}_{m}}}{\tau
					q_{\mathrm{ul},m,j}{{\eta}_{n,m,j}}})}^{2}}}
				{\sum_{j=1}^{{{K}_{m}}}{\tau
						q_{\mathrm{ul},m,j}{{\eta }_{n,m,j}}}
					+\sigma_{\mathrm{ul}}^{2}}}}} \\
\notag	=&\,\sqrt{\sum\limits_{n=1}^{N}{\frac{Lq_{\mathrm{dl},m}\tau
				q_{\mathrm{ul},m,k}\eta_{n,m,k}^{2}}
			{\sum\nolimits_{j=1}^{{{K}_{m}}}
	{\tau q_{\mathrm{ul},m,j}\eta_{n,m,j}}+\sigma_{\mathrm{ul}}^{2}}}} \\
=&\,\sqrt{Lq_{\mathrm{dl},m}{{\zeta }_{m,k}}},
\end{align}
where (b) is obtained by large-scale random matrix conclusions.

\ifCLASSOPTIONcaptionsoff
  \newpage
\fi
\bibliographystyle{IEEEtran}
\bibliography{references}

\begin{thebibliography}{10}
\providecommand{\url}[1]{#1}
\csname url@samestyle\endcsname
\providecommand{\newblock}{\relax}
\providecommand{\bibinfo}[2]{#2}
\providecommand{\BIBentrySTDinterwordspacing}{\spaceskip=0pt\relax}
\providecommand{\BIBentryALTinterwordstretchfactor}{4}
\providecommand{\BIBentryALTinterwordspacing}{\spaceskip=\fontdimen2\font plus
\BIBentryALTinterwordstretchfactor\fontdimen3\font minus
  \fontdimen4\font\relax}
\providecommand{\BIBforeignlanguage}[2]{{%
\expandafter\ifx\csname l@#1\endcsname\relax
\typeout{** WARNING: IEEEtran.bst: No hyphenation pattern has been}%
\typeout{** loaded for the language `#1'. Using the pattern for}%
\typeout{** the default language instead.}%
\else
\language=\csname l@#1\endcsname
\fi
#2}}
\providecommand{\BIBdecl}{\relax}
\BIBdecl

\bibitem{nayebi2015cell}
E.~Nayebi, A.~Ashikhmin, T.~L. Marzetta, and H.~Yang, ``Cell-free massive
  {MIMO} systems,'' in \emph{49th Asilomar Conference on Signals, Systems and
  Computers}, Nov. 2015, pp. 695--699.

\bibitem{buzzi2019user}
S.~Buzzi, C.~D’Andrea, A.~Zappone, and C.~D’Elia, ``User-centric {5G}
  cellular networks: {Resource} allocation and comparison with the cell-free
  massive {MIMO} approach,'' \emph{IEEE Trans. Wireless Commun.}, vol.~19,
  no.~2, pp. 1250--1264, Nov. 2019.

\bibitem{ngo2015cell}
H.~Q. Ngo, A.~Ashikhmin, H.~Yang, E.~G. Larsson, and T.~L. Marzetta,
  ``Cell-free massive {MIMO} uniformly great service for everyone,'' in
  \emph{IEEE 16th international workshop on signal processing advances in
  wireless communications (SPAWC)}, Jun. 2015, pp. 201--205.

\bibitem{li2017downlink}
J.~Li, D.~Wang, P.~Zhu, J.~Wang, and X.~You, ``Downlink spectral efficiency of
  distributed massive {MIMO} systems with linear beamforming under pilot
  contamination,'' \emph{IEEE Transactions on Vehicular Technology}, vol.~67,
  no.~2, pp. 1130--1145, Jul. 2017.

\bibitem{parida2018downlink}
P.~Parida, H.~S. Dhillon, and A.~F. Molisch, ``Downlink performance analysis of
  cell-free massive {MIMO} with finite fronthaul capacity,'' in \emph{IEEE 88th
  Vehicular Technology Conference (VTC-Fall)}, Aug. 2018, pp. 1--6.

\bibitem{riera2019decentralization}
F.~Riera-Palou and G.~Femenias, ``Decentralization issues in cell-free massive
  {MIMO} networks with {Zero-Forcing} precoding,'' in \emph{2019 57th Annual
  Allerton Conference on Communication, Control, and Computing (Allerton)},
  Sept. 2019, pp. 521--527.

\bibitem{yaacoub2016overview}
E.~Yaacoub, M.~Husseini, and H.~Ghaziri, ``An overview of research topics and
  challenges for {5G} massive {MIMO} antennas,'' in \emph{IEEE Middle East
  Conference on Antennas and Propagation (MECAP)}, Dec. 2016, pp. 1--4.

\bibitem{larsson2016joint}
E.~G. Larsson and H.~V. Poor, ``Joint beamforming and broadcasting in massive
  {MIMO},'' \emph{IEEE Transactions on Wireless Communications}, vol.~15,
  no.~4, pp. 3058--3070, Jan. 2016.

\bibitem{zhang2019cell}
J.~Zhang, S.~Chen, Y.~Lin, J.~Zheng, B.~Ai, and L.~Hanzo, ``Cell-free massive
  mimo: A new next-generation paradigm,'' \emph{IEEE Access}, vol.~7, pp.
  99\,878--99\,888, 2019.

\bibitem{ammar2021user}
H.~A. Ammar, R.~Adve, S.~Shahbazpanahi, G.~Boudreau, and K.~V. Srinivas,
  ``User-centric cell-free massive mimo networks: A survey of opportunities,
  challenges and solutions,'' \emph{IEEE Communications Surveys \& Tutorials},
  2021.

\bibitem{wang2021live}
D.~Wang, C.~Zhang, Z.~Ji, Y.~Du, J.~Zhao, M.~Jiang, and X.~You, ``Live
  demonstration: A cloud-based cell-free distributed massive mimo system,'' in
  \emph{2021 IEEE International Symposium on Circuits and Systems
  (ISCAS)}.\hskip 1em plus 0.5em minus 0.4em\relax IEEE, 2021, pp. 1--1.

\bibitem{papazafeiropoulos2020performance}
A.~Papazafeiropoulos, P.~Kourtessis, M.~Di~Renzo, S.~Chatzinotas, and J.~M.
  Senior, ``Performance analysis of cell-free massive mimo systems: A
  stochastic geometry approach,'' \emph{IEEE Transactions on Vehicular
  Technology}, vol.~69, no.~4, pp. 3523--3537, 2020.

\bibitem{liu2019spectral}
P.~Liu, K.~Luo, D.~Chen, and T.~Jiang, ``Spectral efficiency analysis of
  cell-free massive mimo systems with zero-forcing detector,'' \emph{IEEE
  Transactions on Wireless Communications}, vol.~19, no.~2, pp. 795--807, 2019.

\bibitem{femenias2020short}
G.~Femenias, F.~Riera-Palou, A.~{\'A}lvarez-Polegre, and A.~Garc{\'\i}a-Armada,
  ``Short-term power constrained cell-free massive-mimo over spatially
  correlated ricean fading,'' \emph{IEEE Transactions on Vehicular Technology},
  vol.~69, no.~12, pp. 15\,200--15\,215, 2020.

\bibitem{zhang2019closed}
Y.~Zhang and L.~Dai, ``A closed-form approximation for uplink average ergodic
  sum capacity of large-scale multi-user distributed antenna systems,''
  \emph{IEEE Transactions on Vehicular Technology}, vol.~68, no.~2, pp.
  1745--1756, 2019.

\bibitem{hu2017energy}
Y.~Hu, F.~Zhang, C.~Li, Y.~Wang, and R.~Zhao, ``Energy efficiency resource
  allocation in downlink cell-free massive mimo system,'' in \emph{2017
  International Symposium on Intelligent Signal Processing and Communication
  Systems (ISPACS)}.\hskip 1em plus 0.5em minus 0.4em\relax IEEE, 2017, pp.
  878--882.

\bibitem{mosleh2019downlink}
S.~Mosleh, H.~Almosa, E.~Perrins, and L.~Liu, ``Downlink resource allocation in
  cell-free massive mimo systems,'' in \emph{2019 International Conference on
  Computing, Networking and Communications (ICNC)}.\hskip 1em plus 0.5em minus
  0.4em\relax IEEE, 2019, pp. 883--887.

\bibitem{amin2018quantized}
B.~Amin, B.~Abdelhamid, and S.~El-Ramly, ``Quantized power allocation
  algorithms in cell-free massive mimo systems,'' in \emph{2018 International
  Japan-Africa Conference on Electronics, Communications and Computations
  (JAC-ECC)}.\hskip 1em plus 0.5em minus 0.4em\relax IEEE, 2018, pp. 35--38.

\bibitem{riera2019trade}
F.~Riera-Palou and G.~Femenias, ``Trade-offs in cell-free massive mimo
  networks: Precoding, power allocation and scheduling,'' in \emph{2019 14th
  International Conference on Advanced Technologies, Systems and Services in
  Telecommunications (TELSIKS)}.\hskip 1em plus 0.5em minus 0.4em\relax IEEE,
  2019, pp. 158--165.

\bibitem{izadi2020power}
A.~Izadi, S.~M. Razavizadeh, and O.~Saatlou, ``Power allocation for downlink
  training in cell-free massive mimo networks,'' in \emph{2020 10th
  International Symposium onTelecommunications (IST)}.\hskip 1em plus 0.5em
  minus 0.4em\relax IEEE, 2020, pp. 111--115.

\bibitem{chakraborty2020efficient}
S.~Chakraborty, {\"O}.~T. Demir, E.~Bj{\"o}rnson, and P.~Giselsson, ``Efficient
  downlink power allocation algorithms for cell-free massive mimo systems,''
  \emph{IEEE Open Journal of the Communications Society}, vol.~2, pp. 168--186,
  2020.

\bibitem{chen2019dynamic}
Z.~Chen, E.~Bj{\"o}rnson, and E.~G. Larsson, ``Dynamic resource allocation in
  co-located and cell-free massive mimo,'' \emph{IEEE Transactions on Green
  Communications and Networking}, vol.~4, no.~1, pp. 209--220, 2019.

\bibitem{palhares2021robust}
V.~M. Palhares, A.~R. Flores, and R.~C. de~Lamare, ``Robust mmse precoding and
  power allocation for cell-free massive mimo systems,'' \emph{IEEE
  Transactions on Vehicular Technology}, vol.~70, no.~5, pp. 5115--5120, 2021.

\bibitem{ding2021machine}
J.~Ding, D.~Qu, P.~Liu, and J.~Choi, ``Machine learning enabled preamble
  collision resolution in distributed massive mimo,'' \emph{IEEE Transactions
  on Communications}, vol.~69, no.~4, pp. 2317--2330, 2021.

\bibitem{sidiropoulos2006transmit}
N.~D. Sidiropoulos, T.~N. Davidson, and Z.-Q. Luo, ``Transmit beamforming for
  physical-layer multicasting,'' \emph{IEEE transactions on signal processing},
  vol.~54, no.~6, pp. 2239--2251, Jun. 2006.

\bibitem{karipidis2008quality}
E.~Karipidis, N.~D. Sidiropoulos, and Z.-Q. Luo, ``Quality of service and
  max-min fair transmit beamforming to multiple cochannel multicast groups,''
  \emph{IEEE Transactions on Signal Processing}, vol.~56, no.~3, pp.
  1268--1279, Feb. 2008.

\bibitem{yang2013multicast}
H.~Yang, T.~L. Marzetta, and A.~Ashikhmin, ``Multicast performance of
  large-scale antenna systems,'' in \emph{IEEE 14th Workshop on Signal
  Processing Advances in Wireless Communications (SPAWC)}, Sept. 2013, pp.
  604--608.

\bibitem{sadeghi2017reducing}
M.~Sadeghi, L.~Sanguinetti, R.~Couillet, and C.~Yuen, ``Reducing the
  computational complexity of multicasting in large-scale antenna systems,''
  \emph{IEEE Transactions on Wireless Communications}, vol.~16, no.~5, pp.
  2963--2975, Mar. 2017.

\bibitem{christopoulos2015multicast}
D.~Christopoulos, S.~Chatzinotas, and B.~Ottersten, ``Multicast multigroup
  beamforming for per-antenna power constrained large-scale arrays,'' in
  \emph{IEEE 16th International Workshop on Signal Processing Advances in
  Wireless Communications (SPAWC)}, Aug. 2015, pp. 271--275.

\bibitem{dong2020multi}
M.~Dong and Q.~Wang, ``Multi-group multicast beamforming: Optimal structure and
  efficient algorithms,'' \emph{IEEE Transactions on Signal Processing},
  vol.~68, pp. 3738--3753, May 2020.

\bibitem{bandi2020joint}
A.~Bandi, R.~B.~S. Mysore, S.~Chatzinotas, and B.~Ottersten, ``Joint user
  scheduling, and precoding for multicast spectral efficiency in multigroup
  multicast systems,'' in \emph{International Conference on Signal Processing
  and Communications (SPCOM)}, Aug. 2020, pp. 1--5.

\bibitem{sadeghi2018joint}
M.~Sadeghi, E.~Bj{\"o}rnson, E.~G. Larsson, C.~Yuen, and T.~Marzetta, ``Joint
  unicast and multi-group multicast transmission in massive {MIMO} systems,''
  \emph{IEEE Transactions on Wireless Communications}, vol.~17, no.~10, pp.
  6375--6388, 2018.

\bibitem{doan2017performance}
T.~X. Doan, H.~Q. Ngo, T.~Q. Duong, and K.~Tourki, ``On the performance of
  multigroup multicast cell-free massive {MIMO},'' \emph{IEEE Commun. Lett.},
  vol.~21, no.~12, pp. 2642--2645, Aug. 2017.

\bibitem{zhang2019max}
Y.~Zhang, H.~Cao, and L.~Yang, ``Max-min power optimization in multigroup
  multicast cell-free massive {MIMO},'' in \emph{IEEE Wireless Communications
  and Networking Conference (WCNC)}, Apr. 2019, pp. 1--6.

\bibitem{tan2020energy}
F.~Tan, P.~Wu, Y.-C. Wu, and M.~Xia, ``Energy-efficient non-orthogonal
  multicast and unicast transmission of cell-free massive {MIMO} systems with
  {SWIPT},'' \emph{IEEE J. Sel. Areas Commun.}, Sep. 2020.

\bibitem{zhang2019secrecy}
X.~Zhang, D.~Guo, K.~An, Z.~Ding, and B.~Zhang, ``Secrecy analysis and active
  pilot spoofing attack detection for multigroup multicasting cell-free massive
  {MIMO} systems,'' \emph{IEEE Access}, vol.~7, pp. 57\,332--57\,340, Apr.
  2019.

\bibitem{bjornson2015massive}
E.~Bj{\"o}rnson, M.~Matthaiou, and M.~Debbah, ``Massive {MIMO} with non-ideal
  arbitrary arrays: {Hardware} scaling laws and circuit-aware design,''
  \emph{IEEE Transactions on Wireless Communications}, vol.~14, no.~8, pp.
  4353--4368, Apr. 2015.

\bibitem{mohammed2012single}
S.~K. Mohammed and E.~G. Larsson, ``Single-user beamforming in large-scale
  {MISO} systems with per-antenna constant-envelope constraints: The doughnut
  channel,'' \emph{IEEE Transactions on Wireless Communications}, vol.~11,
  no.~11, pp. 3992--4005, Sept. 2012.

\bibitem{yu2016alternating}
X.~Yu, J.-C. Shen, J.~Zhang, and K.~B. Letaief, ``Alternating minimization
  algorithms for hybrid precoding in millimeter wave {MIMO} systems,''
  \emph{IEEE Journal of Selected Topics in Signal Processing}, vol.~10, no.~3,
  pp. 485--500, Feb. 2016.

\bibitem{ngo2013energy}
H.~Q. Ngo, E.~G. Larsson, and T.~L. Marzetta, ``Energy and spectral efficiency
  of very large multiuser {MIMO} systems,'' \emph{IEEE Transactions on
  Communications}, vol.~61, no.~4, pp. 1436--1449, Feb. 2013.

\bibitem{deb2002fast}
K.~Deb, A.~Pratap, S.~Agarwal, and T.~Meyarivan, ``A fast and elitist
  multiobjective genetic algorithm: {NSGA-II},'' \emph{IEEE transactions on
  evolutionary computation}, vol.~6, no.~2, pp. 182--197, Aug. 2002.

\bibitem{WS2018optimization}
W.~S.~A. {Qatab}, M.~Y. {Alias}, and I.~{Ku}, ``Optimization of multi-objective
  resource allocation problem in cognitive radio {LTE/LTE-A} femtocell networks
  using {NSGA II},'' in \emph{Proc. IEEE Int. Symposium on Telecommun. Technol.
  (ISTT)}, Nov 2018, pp. 1--6.

\bibitem{lecun2015deep}
Y.~LeCun, Y.~Bengio, and G.~Hinton, ``Deep learning,'' \emph{nature}, vol. 521,
  no. 7553, pp. 436--444, 2015.

\bibitem{8283585}
K.~{Kim}, J.~{Lee}, and J.~{Choi}, ``Deep learning based pilot allocation
  scheme {(DL-PAS)} for {5G} massive {MIMO} system,'' \emph{IEEE Communications
  Letters}, vol.~22, no.~4, pp. 828--831, Feb. 2018.

\bibitem{wang2019performance}
D.~Wang, M.~Wang, P.~Zhu, J.~Li, J.~Wang, and X.~You, ``Performance of
  network-assisted full-duplex for cell-free massive {MIMO},'' \emph{IEEE
  Transactions on Communications}, vol.~68, no.~3, pp. 1464--1478, Dec. 2019.

\end{thebibliography}
\begin{IEEEbiography}[{\includegraphics[width=1in,height=1.25in,clip,
		keepaspectratio]{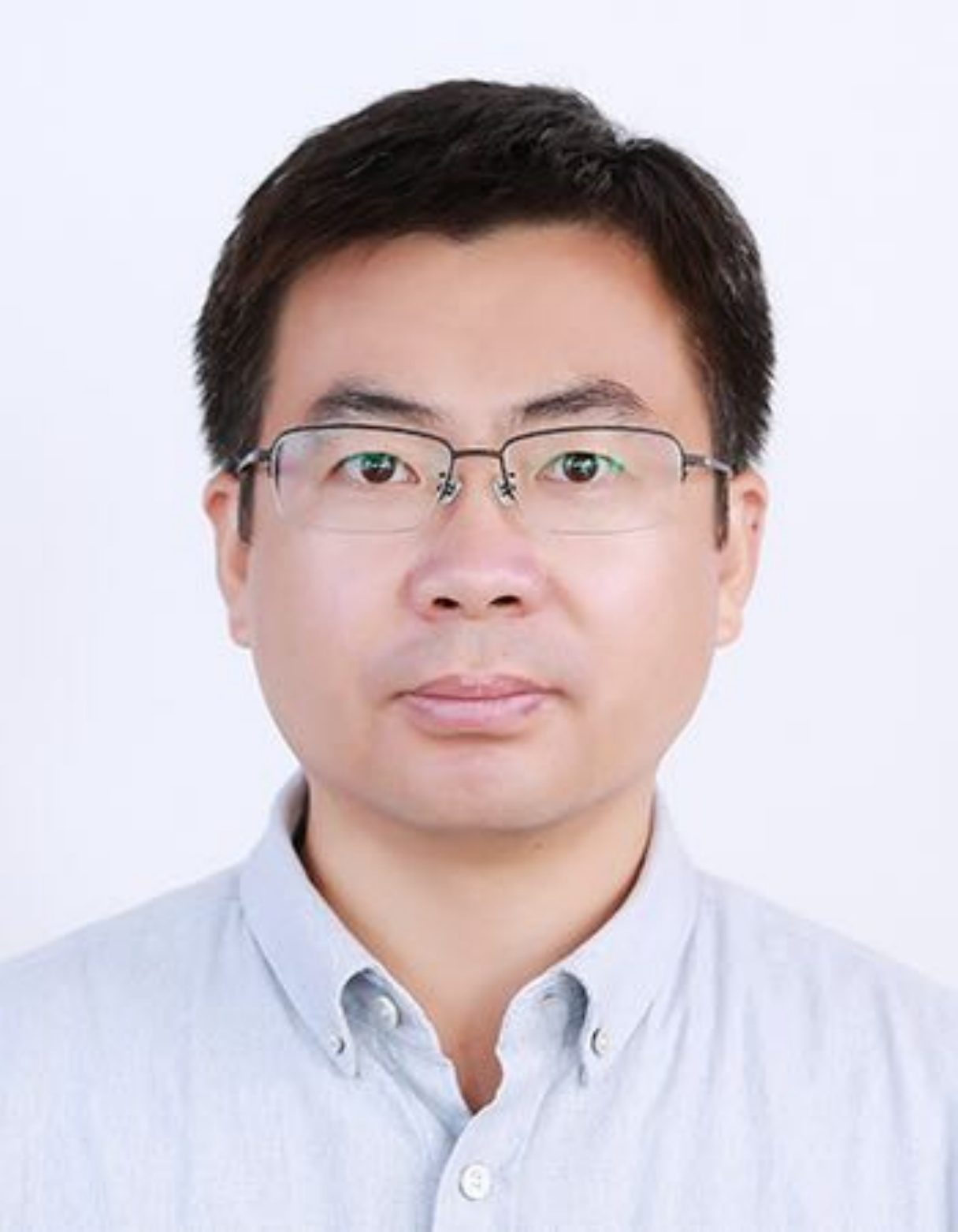}}]{Jiamin
		Li} received the B.S. and M.S. degrees in communication and information 
	systems from Hohai University, Nanjing, China, in 2006 and 2009, 
	respectively, 
	and the Ph.D. degree in information and communication engineering from 
	Southeast University, Nanjing, China, in 2014. He joined the National 
	Mobile 
	Communications Research Laboratory, Southeast University, in 2014, where he 
	has been an Associate Professor since 2019. His research interests include 
	massive MIMO, distributed antenna systems, and multi-objective 
	optimization.\\
\end{IEEEbiography}
\begin{IEEEbiography}[{\includegraphics[width=1in,height=1.30in,clip,
		keepaspectratio]{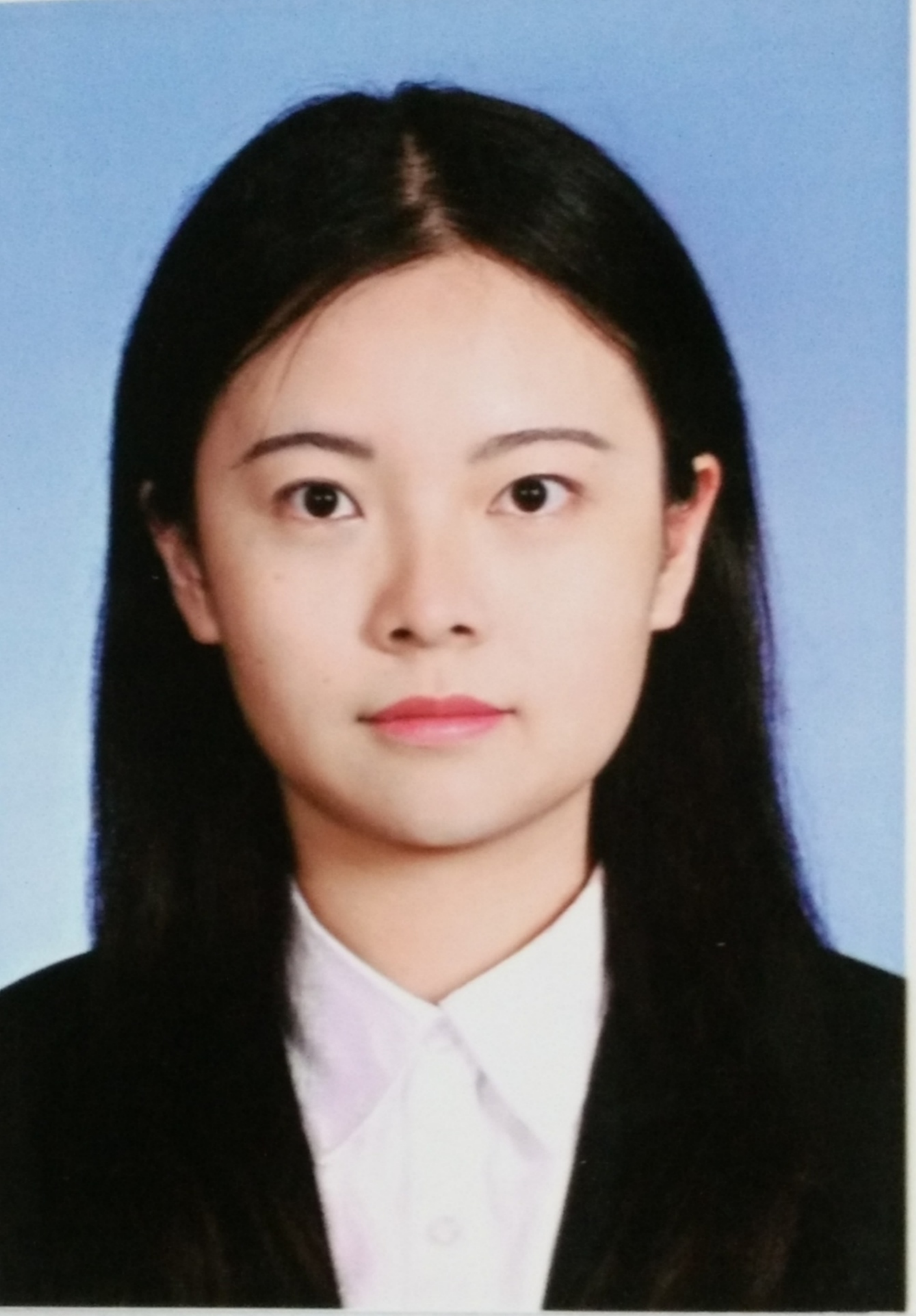}}]{Qijun
		Pan} was born in Shanghai province, China, in 1998. She received the 
		B.S. degree in the school of communication and information engineering 
		from Shanghai University, Shanghai, China, in 2020. She is currently 
		pursuing the M.S. degree in electronic and communication engineering at 
		the National Mobile Communications Research Laboratory, Southeast 
		University. Her research interests include massive MIMO, distributed 
		antenna systems and ultra reliable low latency communication.\\
\end{IEEEbiography}
\begin{IEEEbiography}[{\includegraphics[width=1in,height=1.30in,clip,
		keepaspectratio]{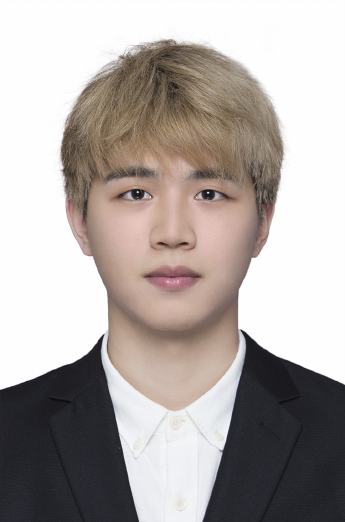}}]{Zhenggang Wu
	} was born in Yangzhou, Jiangsu province, China. He received the B.S. 
	degree in communication engineering from Hohai University, Nanjing, China, 
	in 2020. He is currently working toward the M.S. degree in electronic and 
	communication engineering at the National Mobile Communications Research 
	Laboratory, Southeast University. His research interests mainly include 
	massive MIMO, ultrareliable low latency communication, 3D coverage and 
	massive access.\\
\end{IEEEbiography}
\begin{IEEEbiography}[{\includegraphics[width=1in,height=1.30in,clip,
		keepaspectratio]{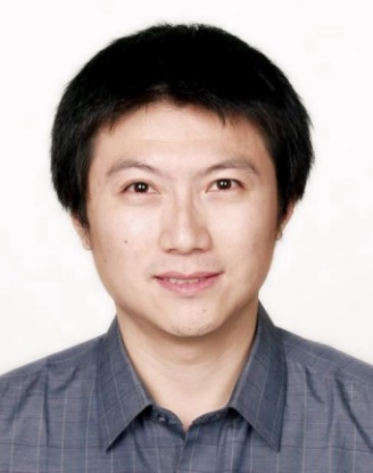}}]{Pengcheng
		Zhu}  received the B.S and M.S. degrees in electrical engineering from 
		Shandong University, Jinan, China, in 2001 and 2004, respectively, and 
		the Ph.D. degree in communication and information science from the 
		Southeast University, Nanjing, China, in 2009. He has been a lecturer 
		with the national mobile communications research laboratory, Southeast 
		University, China, since 2009. His research interests lie in the areas 
		of communication and signal processing, including limited feedback 
		techniques, and distributed antenna systems.\\
\end{IEEEbiography}
\begin{IEEEbiography}[{\includegraphics[width=1in,height=1.30in,clip,
		keepaspectratio]{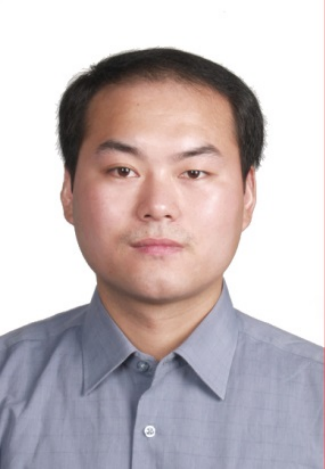}}]{Dongming Wang 
		}  received the B.S. degree from Chongqing University of Posts and 
		Telecommunications, Chongqing, China, the M.S. degree from Nanjing 
		University of Posts and Telecommunications, Nanjing, China, and the 
		Ph.D. degree from the Southeast University, Nanjing, China, in 1999, 
		2002, and 2006, respectively. He joined the National Mobile 
		Communications Research Laboratory, Southeast University, in 2006, 
		where he has been an Associate Professor since 2010. His research 
		interests include turbo detection, channel estimation, distributed 
		antenna systems, and large-scale MIMO systems.\\
\end{IEEEbiography}
\begin{IEEEbiography}[{\includegraphics[width=1in,height=1.30in,clip,
		keepaspectratio]{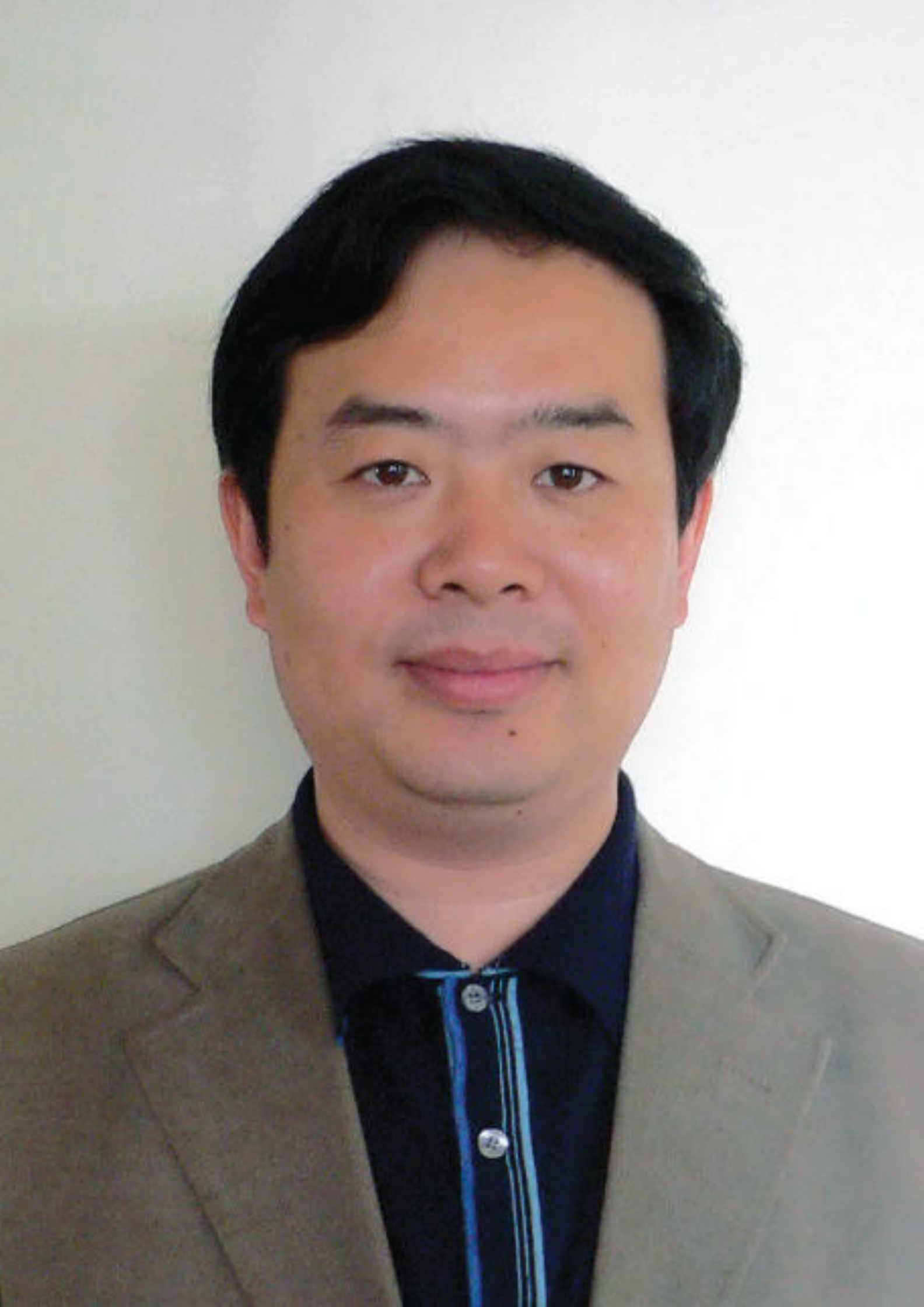}}]{Xiaohu You 
	}   (Fellow, IEEE) received the B.S., M.S. and Ph.D. degrees in electrical 
	engineering from Nanjing Institute of Technology, Nanjing, China, in 1982, 
	1985, and 1989, respectively. From 1987 to 1989, he was with Nanjing 
	Institute of Technology as a Lecturer. From 1990 to the present time, he 
	has been with Southeast University, first as an Associate Professor and 
	later as a Professor. His research interests include mobile communications, 
	adaptive signal processing, and artificial neural networks with 
	applications to communications and biomedical engineering. He is the Chief 
	of the Technical Group of China 3G/B3G Mobile Communication R \& D Project. 
	He received the excellent paper prize from the China Institute of 
	Communications in 1987 and the Elite Outstanding Young Teacher Awards from 
	Southeast University in 1990, 1991, and 1993. He was also a recipient of 
	the 1989 Young Teacher Award of Fok Ying Tung Education Foundation, State 
	Education Commission of China.\\
\end{IEEEbiography}

\end{document}